\documentclass{aa}
\usepackage{graphicx}
\usepackage{amsmath}
\usepackage{txfonts}
\usepackage{lscape}
\usepackage{longtable, lscape}
\def\HII{H\,{\sc{ii}}}

\def\HI{H\,{\sc{i}}}
\def\fs{\hbox{$.\!\!^{\rm s}$}}
\def\fdg{\hbox{$.\!\!^\circ$}}
\def\farcm{\hbox{$.\mkern-4mu^\prime$}}
\def\farcs{\hbox{$.\!\!^{\prime\prime}$}}
\def\arcmin{\hbox{$^\prime$}}
\def\arcsec{\hbox{$^{\prime\prime}$}}

\def\degr{\hbox{$^\circ$}}
\def\h{\hbox{$^{\reset@font\r@mn{h}}$}}
\def\m{\hbox{$^{\reset@font\r@mn{m}}$}}
\def\s{\hbox{$^{\reset@font\r@mn{s}}$}}
\def\msol{\hbox{\kern 0.20em $M_\odot$}}

\def\kms{\hbox{\kern 0.20em km\kern 0.20em s$^{-1}$}}
\def\cmmt{\hbox{\kern 0.20em cm$^{-3}$}}
\def\cmmd{\hbox{\kern 0.20em cm$^{-2}$}}
\def\pc{\hbox{\kern 0.20em pc$^{2}$}}

\def\h13cop{\hbox{H$^{13}$CO$^{+}$}}

\begin{document}
   \title{Triggered star formation \\
on the borders of the Galactic \HII\ region RCW~82}

   \author{M.~Pomar\`es\inst{1}
            \and
            A.~Zavagno\inst{1}
            \and
            L.~Deharveng\inst{1}
            \and
            M.~Cunningham\inst{2}
            \and
            P.~Jones\inst{2}
           \and
           S.~Kurtz\inst{3}
          \and
         D.~Russeil\inst{1}
          \and
         J.~Caplan\inst{1}
         \and
         F. Comer\'on\inst{4}
          }

   \offprints{M. Pomar\`es}

   \institute{Laboratoire d'Astrophysique de Marseille / UMR6110, CNRS / Universit\'e de Provence, Technop\^ole de
   Marseille-Etoile, 38 rue Fr\'ed\'eric Joliot-Curie, 13388 Marseille CEDEX 13, France
         \and School of Physics, University of New South Wales, Sydney, NSW 2052, Australia
         \and Centro de Radioastronom\'ia y Astrof\'isica, Universidad Nacional Aut\`onoma de Mexico, Apartado Postal 3-72, 58089 Morelia, Michoac\'an, M\'exico
         \and European Southern Observatory, Karl-Schwarzschild-Strasse 2, 85748 Garching, Germany\\
         \email{melanie.pomares@oamp.fr}
             }

   \date{Received ; accepted }

\abstract
{We are engaged in a multi-wavelength study of several Galactic
\HII\ regions that exhibit signposts of triggered star formation
on their borders, where the collect and collapse process could be
at work.}
{When addressing the question of triggered star formation, it is
critical to ensure the {\emph{real}} association between the
ionized gas and the neutral material observed nearby. In this
paper we stress this point and present CO observations of the
RCW~82 star forming region.}
{The velocity distribution of the molecular gas is combined with
the study of young stellar objects (YSOs) detected in the
direction of RCW~82. We discuss the YSO's evolutionary status
using near- and mid-IR data. The spatial and velocity
distributions of the molecular gas are used to discuss the
possible scenarios for the star formation around RCW~82.}
{Several massive molecular condensations, together with star
formation sites, are observed on the borders of RCW~82.  The
shapes of the three brightest condensations suggest that they were
pre-existent, i.e. not formed through the collect and collapse
process. A thin layer of molecular material is observed
surrounding the ionized gas, adjacent to the ionization front.
This results from the sweeping up of neutral material around the
expanding region. Several Class~I YSOs are detected in the
direction of this layer.}
{The numerous YSOs observed towards the bright molecular
condensations bordering (and velocity-associated with) the ionized
gas reveal the intense star formation activity in RCW~82. But this
region is probably too young to have triggered star formation via
the collect and collapse process.}

 \titlerunning{Triggered star formation on the borders of RCW~82}
 \maketitle

\section{Introduction \label{intro}}

The formation of massive stars can be triggered by \HII\ regions.
Massive molecular condensations on the borders of Galactic \HII\
regions are the most likely sites for star formation and are
consequently the most likely locations for observing the earliest
stages of stellar births.

Several recent \HII\ region studies (Deharveng et
al.~\cite{deh03}, Zavagno et al.~\cite{zav06}, Watson et
al.~\cite{Wat08}, Cappa et al.~\cite{cap08}, Kirsanova et
al.~\cite{kir08}) have taken up the challenge to understand the
triggered star formation processes, even if some points still
remain not clear.

In the collect and collapse process, proposed by Elmegreen \& Lada
(\cite{elm77}), the expansion of the ionized region leads to the
formation of a dense layer of material, collected between the
ionization front (IF) and the shock front, during the
supersonic expansion of the \HII\ region. This layer eventually
becomes gravitationally unstable along its length and fragments,
leading to the formation of massive condensations that are
potential sites for subsequent star formation. In 2005, Deharveng
et al.\ proposed a list of candidate Galactic \HII\ regions where
the collect and collapse process may be the main triggering agent
for the star formation observed on their borders. We have shown
this process to be at work in a number of regions, among them
Sh~104 (Deharveng et al. \cite{deh03}) and RCW~79 (Zavagno et
al.~\cite{zav06}). We have also shown that this process works in
inhomogeneous turbulent clouds as in the cases of Sh2-212
(Deharveng et al.~\cite{deh08}), Sh2-217 (Brand et al. in
preparation) and Sh2-241 (Pomar\`es et al. in preparation).

The observations of an increasing number of sources show that the
layer of collected neutral material is always present around the
ionized gas and that various triggering processes are at work
there.

Information about the distribution of associated molecular
material is frequently lacking for these regions. We therefore
carried out line observations of the molecular material, at
velocities bracketing those of the surrounding ionized gas.
Without such radial velocities we are never sure that a given
condensation is indeed associated with the \HII\ region we are
studying. The same is true of the YSOs that are observed {\emph{in
the direction of}} the region, with no more than a reasonable
probability of being associated with it, based on the spatial
distribution of these types of sources. We therefore carried out a
series of molecular observations around \HII\ regions using the
Mopra radio-telescope. The first results are presented here for
RCW~82, along with a discussion of the YSO population using new
near-IR, Spitzer-GLIMPSE and MIPSGAL mid-IR data.

This paper is organized as follows: Sect.~2 presents the Galactic
\HII\ region RCW~82. New observations towards this region are
presented in Sect.~3. Results are given in Sect.~4. Sect.~5
presents a general discussion about the morphology of the region
and the star formation processes at work there. Conclusions are
given in Sect.~6.


\section{Presentation of RCW~82 \label{presentation}}

RCW~82 (Rodgers et al.~\cite{rod60}) is an optical \HII\ region
centred at $\alpha_{2000}$=13$^{\rm h}$ 59$^{\rm m}$ 30$^{\rm
s}$, $\delta_{2000}$=$-$61$\degr$ 23$\arcmin$ 30$\arcsec$
($l$=310\fdg99, $b$=0\fdg41). Its diameter is $\sim$6\arcmin\
($\sim$6~pc for a distance of 3.4~kpc).

RCW~82 is a thermal radio-continuum source (Whiteoak et
al.~\cite{whi94}). Its radio flux density is very badly
constrained from the literature: $S_{\rm{5~GHz}}=0.73$~Jy (Haynes
et al.~\cite{hay79}), $S_{\rm{5~GHz}}=4.6$~Jy (Caswell \& Haynes
\cite{cas87}; resolution 4\farcm2), $S_{\rm{4.85~GHz}}=1.302$~Jy
(Kuchar et al.~\cite{kuc97}), $S_{\rm{843~MHz}}=1.5$~Jy (Cohen et
al.~\cite{coh07}; resolution 43\arcsec).

\begin{figure}
\includegraphics[width=90mm]{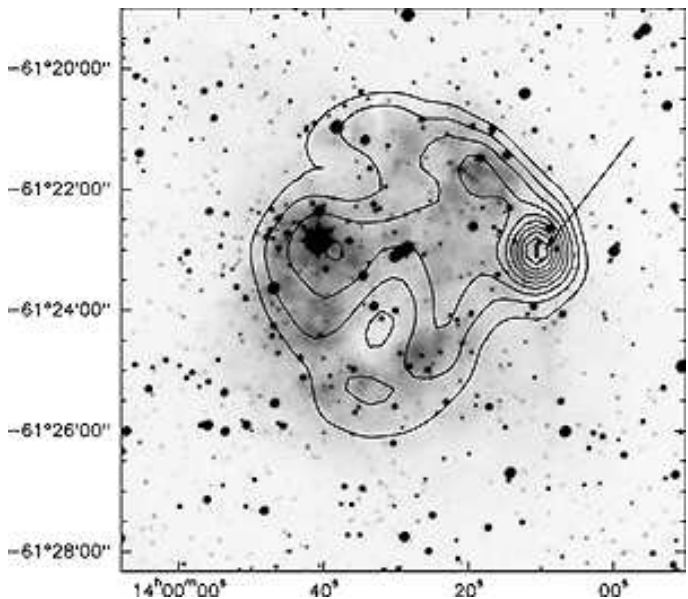}
\caption{SuperCOSMOS image of RCW~82, showing the H$\alpha$
emission of the ionized gas. The contours are for the SUMSS survey
at 843~MHz (FWHM $43\arcsec\times43\arcsec\cos(\delta)$). The
first contour is at 0.02~Jy/beam, and the step is 0.01~Jy/beam.
The arrow points to the radio-continuum point source discussed in
the text. } \label{Haradio}
\end{figure}

Fig.~\ref{Haradio} shows the emission from the ionized gas. The
H$\alpha$ emission (gray scale image) is from the SuperCOSMOS
survey (Parker et al.~\cite{par05}). The contours are for the
SUMSS survey at 843~MHz (FWHM
$43\arcsec\times43\arcsec\cos(\delta)$; Bock, Large, \& Sadler
\cite{boc99}). A radio point source is present in the direction of
the ionized gas, at $\alpha_{2000}$=13$^{\rm h}$ 59$^{\rm m}$
10$^{\rm s}.72$, $\delta_{2000}$=$-$61$\degr$ 23$\arcmin$
00$\arcsec$. At 843~MHz the emission is strongly weighted by
non-thermal emission, if any is present. It is not clear if this
point source is thermal, or if it is a non-thermal background
source, possibly a galaxy (its flux density at 843~MHz is $\sim
70$~mJy, M. Cohen, private communication).

\subsection{RCW~82 in the infrared}

\begin{figure}[htp]
 \includegraphics[width=90mm]{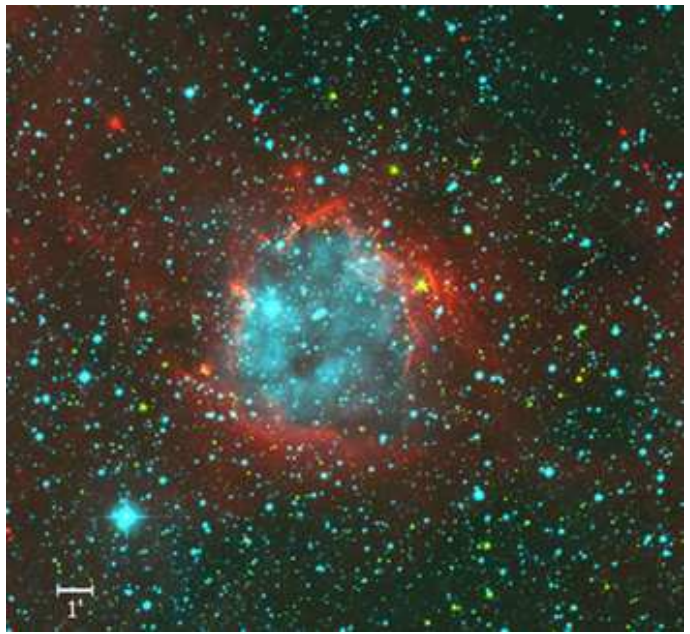}
  \caption{Colour composite image of RCW~82. Spitzer-IRAC 8.0~$\mu$m
  (red) and 3.6~$\mu$m (green) images from the Spitzer-GLIMPSE survey have
  been combined with the H$\alpha$ (blue) image from the SuperCOSMOS
  survey. A PAH emission layer, emitted by the hot
  photo-dissociation region, surrounds the ionized region. The field
  size is 20$\farcm$0 (E-W) $\times$ 18$\farcm$5 (N-S). North is
  up and east is left.}
  \label{I4I1Ha}
\end{figure}

RCW~82 was observed by the infrared satellite MSX (Mid-course
Space eXperiment), from 8.3~$\mu$m to 21.3~$\mu$m (Price et
al.~\cite{pri01}). Deharveng et al.~(\cite{deh05}) present a
colour composite image of RCW~82 (their Fig.~1). At 8.3~$\mu$m,
RCW~82 appears as an empty bubble. The MSX band at 8.3~$\mu$m
contains, superimposed on a continuum, emission bands attributed
to large molecules such as polycyclic aromatic hydrocarbons (PAHs;
L\'eger \& Puget \cite{leg84}). These molecules are destroyed
inside the ionized gas, but are excited in the photodissociation
region (PDR) by the radiation leaking from the H\,{\sc{ii}}
region. On the other hand, part of the 21.3~$\mu$m emission comes
from small grains within the ionized gas of the bubble. Small
grains can survive in strong UV fields, while large molecules
cannot (Cr\'et\'e et al.~\cite{cre99}, Peeters et
al.~\cite{pee02}). RCW~82 (S~137 in Churchwell et al.'s catalogue,
\cite{chu06}) is one of the many bubbles detected by the Spitzer
infrared satellite in the Galactic plane. Fig.~\ref{I4I1Ha} is a
colour composite image of RCW~82, showing the H$\alpha$ emission
of the ionized gas (from the SuperCOSMOS survey) in blue, the
emission at 3.6~$\mu$m in green (mainly stellar emission), from
the Spitzer-GLIMPSE survey (Benjamin et al.~\cite{ben03}), and the
PAH emission at 8.0~$\mu$m in red (Spitzer-GLIMPSE survey). This
confirms, with higher spatial resolution and sensitivity, the view
given by MSX. The PAH emission comes from the hot
photo-dissociation zone surrounding the ionized region. This zone
is extended, but the brightest emission comes from a thin layer
adjacent to the ionization front.

\begin{figure}
 \includegraphics[width=90mm ]{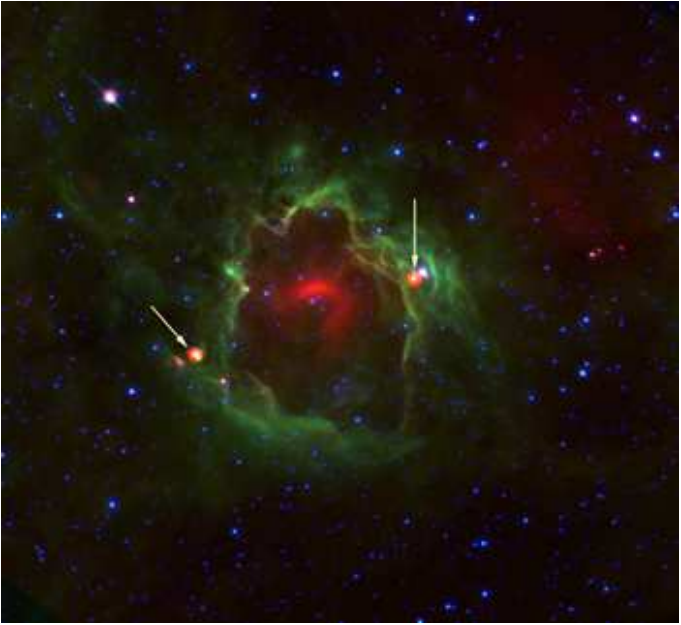}
  \caption{Colour composite image of RCW~82. Red is the emission
of small grains at 24~$\mu$m, from the Spitzer-MIPSGAL survey,
green is the PAH emission at 8.0~$\mu$m, and blue the 4.5~$\mu$m
emission, from the Spitzer-GLIMPSE survey. North is up, east is
left. The arrows point to the two main star formation sites
discussed in Sect~5.2.2. The east and west arrows point to the
cluster and to the YSO $\#$106, respectively. }
  \label{fig_M1I4I2}
\end{figure}

The Spitzer-MIPSGAL emission at 24~$\mu$m (Carey et
al.~\cite{car05}) is dominated by the continuum emission of small
grains, located either in the extended ionized region and its PDR,
or in the envelopes or disks associated with YSOs. This is
illustrated by Fig.~\ref{fig_M1I4I2}, a colour composite image of
RCW~82, showing the stellar emission at 4.5~$\mu$m (in blue), the
PAH emission at 8.0~$\mu$m (in green), and the 24~$\mu$m emission
of the small grains (in red). YSOs and red giants are strong
emitters at 24~$\mu$m. Two bright YSOs (indicated by arrows on
Fig.~\ref{fig_M1I4I2}) are diametrically opposite each other on
the border of RCW~82; they will be discussed in Sect.~5.2.2. We
also see the extended emission of the small grains located inside
the \HII\ region. However, their distribution differs from that of
the ionized gas: {\it i)} they are located in the very central
zone; this may be due to a higher temperature of the grains close
to the central exciting star(s); {\it ii)} their emission is not
enhanced in the direction of the radio-continuum point source
(arrow in Fig.~1); this suggests that this radio point source is
an unrelated non-thermal background source.

\begin{figure}
 \includegraphics[width=90mm]{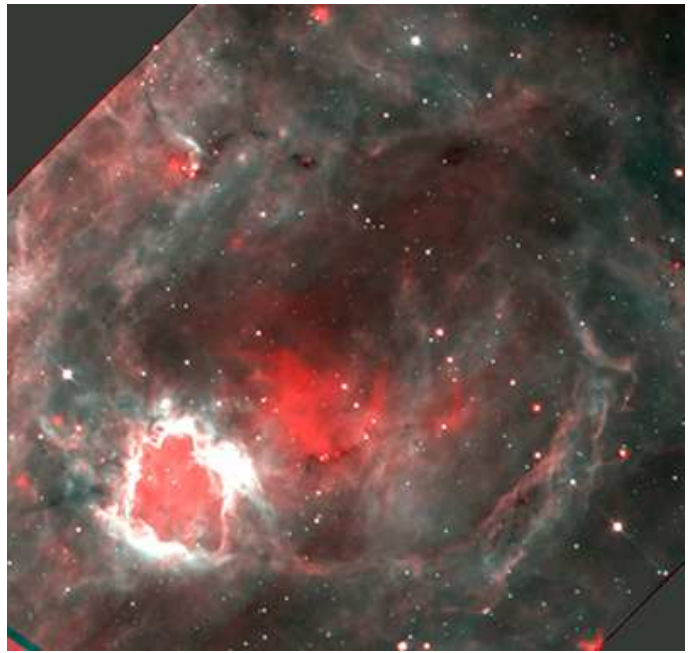}
  \caption{Large scale composite colour image of RCW~82 and its
  surroundings. The 8~$\mu$m emission from Spitzer-GLIMPSE appears in turquoise and the
  24~$\mu$m emission from Spitzer-MIPSGAL in red. North is up, east is left.}
  \label{bulle_grandeEchelle}
\end{figure}

RCW~82 is situated on the southeastern border of a $\sim
27\arcmin$ diameter shell. Fig.~\ref{bulle_grandeEchelle} shows a
colour composite image of this shell, with 8.0~$\mu$m emission in
turquoise and 24~$\mu$m  emission in red. Diffuse radio emission
at 843~MHz is detected by the SUMSS in the direction of the shell
interior, suggesting the presence of ionized gas within the shell.
Most probably the shell surrounds an \HII\ region, and the
24~$\mu$m emission arise from small dust grains surviving with the
ionized gas, as it is the case for RCW~82. We also see several
very luminous red sources in the direction of the shell. They are
associated with dark filaments seen in absorption at 8~$\mu$m,
probably tracing very dense material.
Fig.~\ref{bulle_grandeEchelle} suggests that the formation of
RCW~82 could have been triggered by the expansion of the large
shell. However, we do not know the shell's distance, and whether
it is really associated with RCW~82. This shell is faint and is
not detected on the IRAS survey. We searched the \HI\
(continuum and line) SGPS data (McClure-Griffiths et
al.~\cite{mcc01}) to identify this structure. Continuum emission
at 21~cm confirms that ionized gas is observed towards the
interior of the shell, as well as in RCW~82. However, 21~cm line
emission is observed everywhere and the large shell is not clearly
outlined. Large scale, higher resolution data are needed to
address the association of RCW~82 with this large bubble.

\subsection{The exciting star(s) of RCW~82}

No exciting star of RCW~82 was identified in the literature.
Here we offer indirect evidence suggesting the possible location
and properties of the exciting star(s).

The first piece of information is given by the radio continuum
emission of RCW~82. We used equation (1) of Simpson \& Rubin
(\cite{sim90}) to estimate the number of ionizing Lyman continuum
photons $N_{\rm{Lyc}}$ emitted by the exciting star(s) from the
radio-continuum flux density; Using the radio flux
$S_{5~\rm{GHz}}=4.6$~Jy (Caswell \& Haynes \cite{cas87}),
supposing an electron temperature $T_{\rm{e}}=10000$~K and a
distance of 3.4~kpc, we obtain $\log(N_{\rm{Lyc}}$)$\sim$48.7
photons per second. According to the calibration of O stars by
Martins, Schaerer \& Hillier (\cite{mar05}), this corresponds to a
spectral type $\sim$O7V (assuming a single main-sequence exciting
star).

The second indication concerns the location of the exciting
star(s). Four stars lying near the upper centre of RCW~82 are
candidates. Our selection is based on the following reasons:

\noindent $\bullet$ The \HII\ region is almost circular. Hence the
exciting star(s) are expected near its centre.

\begin{figure}
 \includegraphics[width=90mm ]{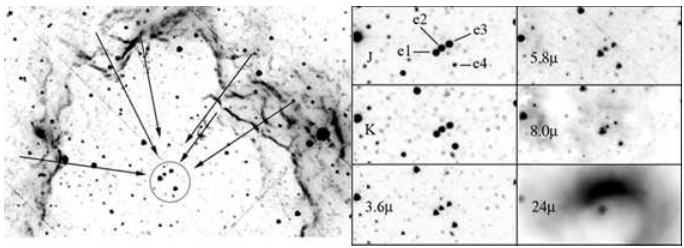}
  \caption{The four candidate exciting stars, e1, e2, e3, and e4. {\it Left:}
  unsharp masked image at 5.8~$\mu$m (from Spitzer-GLIMPSE)
  showing bright rims and ``fingers"  pointing to the candidate exciting stars. {\it Right:} the four central stars at
  various wavelengths.}
  \label{Ex_Stars}
\end{figure}

\noindent $\bullet$ The IF shows some structures (like ``pillars''
or ``fingers'') protruding inside the ionized region
(Fig.~\ref{Ex_Stars}). Such structures, shaped by the UV radiation
field, usually point towards the exciting stars
(Osterbrock~\cite{ost57}). Here, they point to the four central
stars.

\noindent $\bullet$ Another argument is the presence of a hole in
the 24~$\mu$m emission in the centre of RCW~82, seen in the
direction of these four stars, and the annular shape of the dust
emission delimiting its border, clearly seen in
Fig.~\ref{fig_M1I4I2}. This feature is observed in several \HII\
regions such as RCW~120 (Zavagno et al.~\cite{zav07}) and N~49
(Watson et al.~\cite{Wat08}) around the exciting stars.

Fig.~\ref{Ex_Stars} shows these stars, labelled from e1 to e4, at
different wavelengths, from the near-IR to 24~$\mu$m. These stars
have similar $K$-band fluxes, but e2 and e3 decrease in brightness
with increasing wavelength, while e4 is brightest in $K$. Also e1
is brighter than e2 and e3 in the mid-IR; it is the only star
detected at 24~$\mu$m.

We used the 2MASS and Spitzer-GLIMPSE magnitudes to determine the
nature of these four stars, and especially whether they could be O
stars. The 2MASS $J-H$ vs. $H-K$ diagram (Fig.~\ref{cc_JHK}) shows
that e4 is probably a giant, that e1 presents a small near-IR
excess, and that e2 and e3 are probably main sequence stars. The
$J$ vs. $J-K$ diagram (not shown here) confirms that e4 is a
giant, and that stars e2 and e3 are O type stars -- possibly O7.5V
and O6.5V stars with a visual extinction $\sim$ 4.2~mag and
4.7~mag, respectively. In the Spitzer $[3.6]-[4.5]$ vs.
$[5.8]-[8.0]$ colour-colour diagram (Fig~\ref{cc_Spitzer}) star e1
again shows an infrared excess (it appears close to the Class~II
YSO area of the diagram), whereas stars e2, e3 and e4 are situated
in the group of main-sequence and giant stars. Thus we propose
that stars e2 and e3 participate in the excitation of the \HII\
region; the part played by star e1, possibly still accreting
material through a disk, is uncertain. Star e4, a giant, is
possibly unassociated with RCW~82.

According to Martins, Schaerer \& Hillier (\cite{mar05}) the
ionizing flux of an O7.5V plus an 06.5V stars is $\sim 9\times
10^{48}~s^{-1}$, thus slightly larger than necessary to account
for the observed radio flux density of RCW~82. Some of these
ionizing photons are possibly used to heat the dust emitting at
24~$\mu$m in the ionized gas.

\section{New observations and data reduction}

\subsection{Molecular Observations}

We used the 22-m Mopra single-dish radio telescope to study the
distribution of the molecular material associated with RCW~82.
This antenna has a full width at half maximum (FWHM) of
33\arcsec$\pm$2\arcsec and a main beam efficiency of 0.42 at a
frequency of 115~GHz (Ladd et al.~\cite{lad05}). Observations were
undertaken in on-the-fly (OTF) mapping mode using the Mopra-MOPS
digital filterbank (MOPS-DFB) spectrometre in July and August
2007. The MOPS-DFB has a bandwidth of 8~GHz, and can operate in
either broadband or zoom mode (see Jones et al.~\cite{jon08} for
more information). We used the zoom mode of the MOPS Spectrometre,
with 12 'zoom' windows of width 137~MHz, with 4096 channels each,
giving a velocity resolution of 0.09~km~s$^{-1}$. The pointing was
done using nearby SiO masers, with the pointing being generally
better than 10$\arcsec$. Individual maps of
5$\arcmin$~$\times$~5$\arcmin$ were combined to cover a field of
12\farcm5~$\times$~12\farcm5 centred on RCW~82. The OTF maps were
made in a similar way to that described in Bains et al.
(\cite{bai06}). Position switching was used for calibration, with
the offset reference position l=310.6, b=+1.5 being observed after
each row of an OTF map. The scan rate was 2$\arcsec$ per second,
and the spectra were read out with 2 seconds of integration time.
The scan lines were separated by about 10$\arcsec$. Each map took
85 mins to complete, which included 72 mins of on-source time.
Each 5$\arcmin$~$\times$~5$\arcmin$ region was mapped twice, once
scanning in right ascension, and once scanning in declination.
Several lines were observed simultaneously; the lines detected
were $^{12}$CO(1-0) (115.271~GHz), $^{13}$CO(1-0) (110.201~GHz)
and C$^{18}$O(1-0) (109.782~GHz).

Data reduction was done with the ATNF GRIDZILLA and
LIVEDATA packages (http://www.atnf.csiro.au/computing/software/).
Using LIVEDATA, we fitted a linear baseline, computed using
emission-free channels, and subtracted it from each
5$\arcmin~\times~5\arcmin$ map. We then used GRIDZILLA with a
0$\farcm$25 FWHM Gaussian smoothing with a pixel size of
0$\farcm$2 to grid the spectra into datacubes. In this way we
obtained data cubes covering the velocity
range $-100$ to $100$~km~s$^{-1}$ for each spectral line.

\subsection{Near-IR observations}

NTT-SofI observations were obtained at La Silla (Chile) on 15
February 2003, in the $J$ (1.247~$\mu$m), $H$ (1.653~$\mu$m) and
$K_{\rm{S}}$ (2.162~$\mu$m) bands.

Two fields of $\sim$ 2.6$\arcmin$ $\times$ 2.6$\arcmin$ were
observed. Field~1 is centred on the MSX point source
G331.0341+00.3791 centred at $\alpha_{2000}$=13$^{\rm h}$
59$^{\rm m}$ 57$\fs$4, $\delta_{2000}$=$-$61$\degr$ 24$\arcmin$
32$\farcs$9 (see Fig~\ref{fig_M1I4I2}). The position of this MSX
source is indicated in the paper Deharveng et al.~(\cite{deh05}),
where the numbers of the sources in their Fig.~18 are reversed by
mistake. Field~2 is centred at $\alpha_{2000}$=13$^{\rm h}$
59$^{\rm m}$ 45$\fs$6, $\delta_{2000}$=$-$61$\degr$ 22$\arcmin$
54$\farcs$5, on the ionization front.

The images were reduced with the procedure described by Zavagno et
al.~(\cite{zav06}) for the RCW~79 \HII\ region. For field 1, the
total exposure times are 250~s, 36~s and 12~s respectively in $J$,
$H$ and $K_{\rm{S}}$. For field 2, they were 100~s in $J$, 60~s in
$H$ and 24~s in $K_{\rm{S}}$. The photometry was performed using
DAOPHOT (Stetson~\cite{ste87}), also as described in Zavagno et
al.~(\cite{zav06}).

\section{Results}

\subsection{H$\alpha$ velocity field}

\begin{figure}
 \includegraphics[width=90mm ]{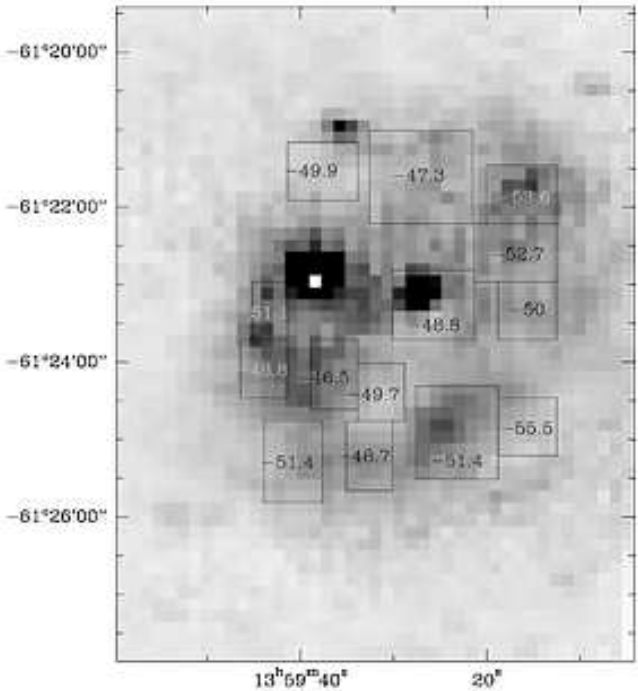}
  \caption{H$\alpha$ Velocity field of RCW~82. The velocities obtained over the selected areas are
  shown. The values in white correspond to the regions where small line width is observed.}
  \label{Fig.delphine_ProfileHalpha}
\end{figure}

Large scale H$\alpha$ emission from the ionized gas around RCW~82
was reported by Russeil et al. (\cite{rus98}). They showed that in
addition to the local diffuse emission, widely distributed diffuse
emission is detected at velocities around $-$31~km~s$^{-1}$ and
$-$50~km~s$^{-1}$. Using the stellar distance of the OB stars and
clusters in the ``313$\degr$ zone" one can place the
$-$31~km~s$^{-1}$ layer at 1.5~$\pm$~0.3~kpc and the
$-$50~km~s$^{-1}$ layer at 3.4~$\pm$~0.9~kpc (Russeil et
al.~\cite{rus98}). With an H$\alpha$ systemic velocity of
$-$50~km~s$^{-1}$ (obtained from the profile integrated over the
total \HII\ region) RCW~82 clearly belongs to the more distant
layer. From these same data we can carry out a detailed study of
the internal kinematics of RCW~82
(Fig.~\ref{Fig.delphine_ProfileHalpha}). The FWHM line width of
the fitted gaussian components has a typical value, between 23 and
25~km~s$^{-1}$, except at three positions where it is less than
20~km~s$^{-1}$. These positions are located on the border of the
ionized region.

\subsection{The molecular materiel}

\subsubsection{Spatial distribution of the molecular materiel\label{sect.Descrption_Mol_Material}}

\begin{figure}
 \includegraphics[width=90mm ]{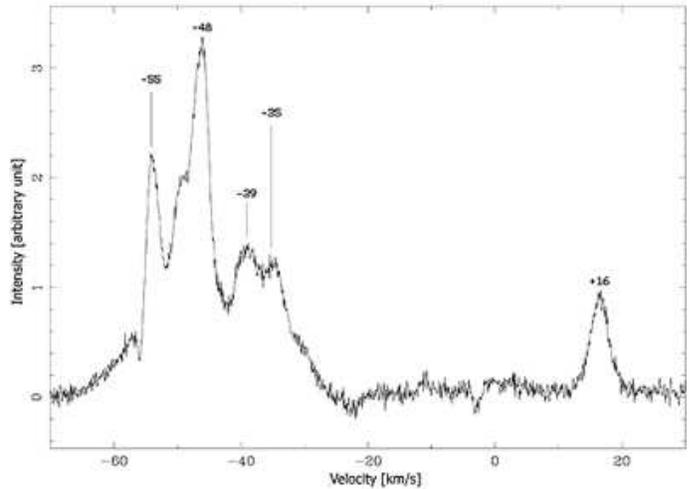}
  \caption{Integrated emission of $^{12}$CO (linear arbitrary unit) on the RCW~82 line of sight. Velocities are
  indicated for the different emission components mentioned in the text.}
  \label{Fig.profile_vitesse_12CO}
\end{figure}

We used the $^{12}$CO, $^{13}$CO and C$^{18}$O datacubes to
study the dynamics of the molecular material detected in the
direction of RCW~82 and to determine some physical properties of
the observed condensations. The first step is to isolate the
molecular components along the line of sight that are associated
with the \HII\ region. Fig.~\ref{Fig.profile_vitesse_12CO} shows
the velocity profile of the $^{12}$CO emission, integrated over
the whole field, between $-70$ and $+30$~km~s$^{-1}$ (no CO
emission is seen outside this velocity range), obtained using
MIRIAD (Sault et al.~\cite{sau95}). We detect at least five
velocity components along the line of sight. The brightest ones
are centred at $-48$ and $-55$~km~s$^{-1}$. Two intermediate ones
are seen at about $-35$ and $-39$~km~s$^{-1}$. Another component
is detected a a positive velocity, $+16$~km~s$^{-1}$.

\begin{figure*}
 \includegraphics[width=160mm]{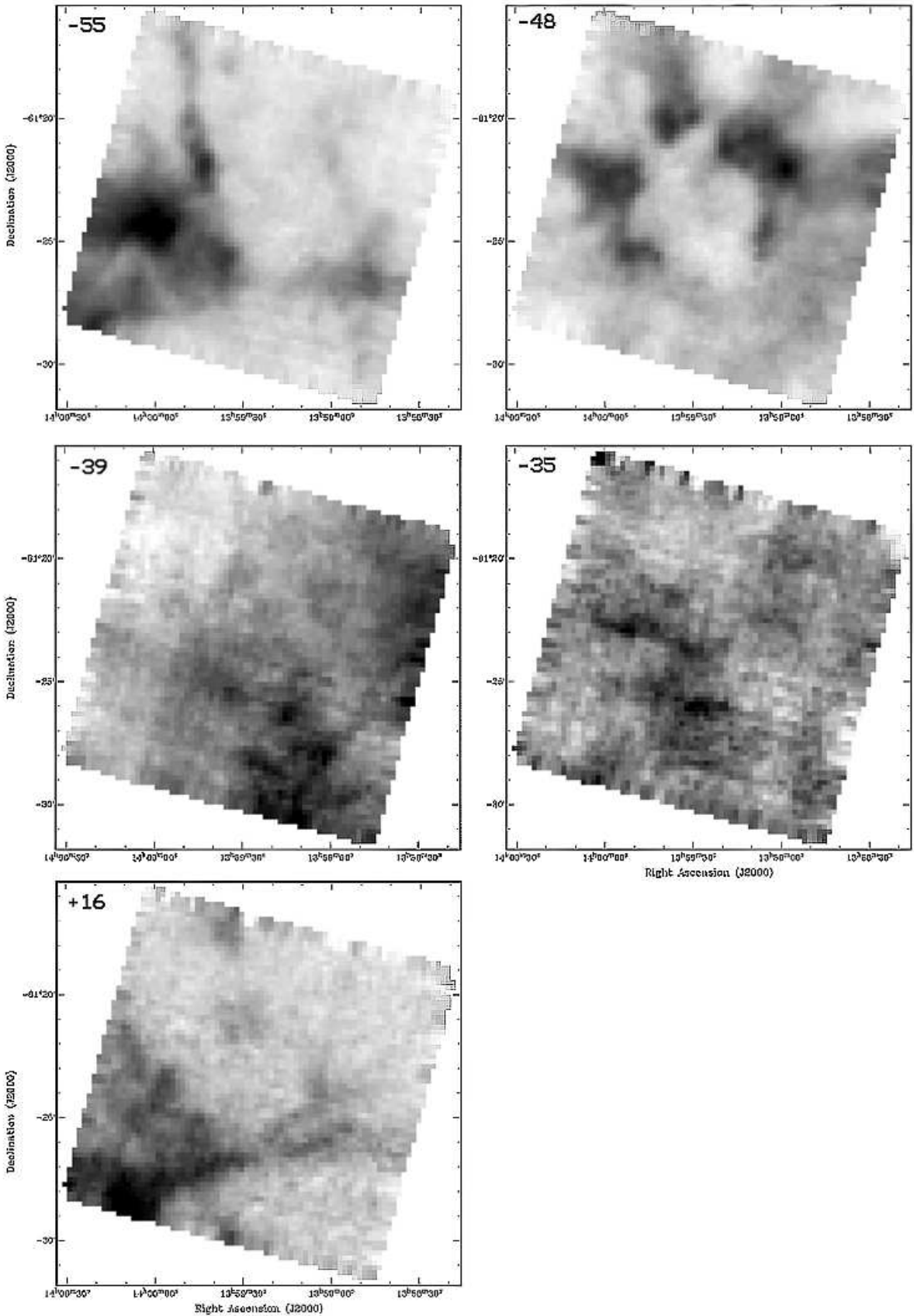}
  \caption{The different components seen in $^{12}$CO on the line of sight
  toward RCW~82. The velocity ranges for each integration are respectively
  $-57.06$ to $-52.07$, $-52.07$ to $-41.99$, $-41.99$ to $-36.35$,
  $-36.35$ to $-23.06$ and $+10.06$ to $+22.07$~km~s$^{-1}$. The central
  velocity is given at the top left of each panel, in km~s$^{-1}$. The
  gray scale is in units of K~km~s$^{-1}$, with darker shades indicating
  higher values.}
  \label{Fig.pannel_lignedevise_12CO}
\end{figure*}

We isolated these 5 molecular components.
Fig.~\ref{Fig.pannel_lignedevise_12CO} shows the morphology of
the corresponding emitting regions.

$\bullet$ The $-55$~km~s$^{-1}$ component is seen as a strong
emission, mainly on the east side of the \HII\ region and partly
on its southwest side. At this velocity, the molecular material
may be interacting with RCW~82, since H$\alpha$ emission from the
\HII\ region is seen at a similar velocity
(Fig.~\ref{Fig.delphine_ProfileHalpha}). The V-shape filamentary
molecular emission associated with the $-55$~km~s$^{-1}$ component
(see Fig.~\ref{Fig.pannel_lignedevise_12CO}) follows the 8~$\mu$m
emission associated with the large bubble
(Fig.~\ref{bulle_grandeEchelle}). This association could indicate
dynamical interactions between this large bubble and RCW~82.
Unfortunately our spatial coverage of Mopra observations is not
sufficient to make this association more clear. The dense emission
peak (see Fig.~\ref{Fig.pannel_lignedevise_12CO} and
Fig.~\ref{Fig.struct-55}) coincides with a dark structure seen at
8~$\mu$m (Fig.~\ref{bulle_grandeEchelle}).

\begin{figure}
 \includegraphics[width=90mm]{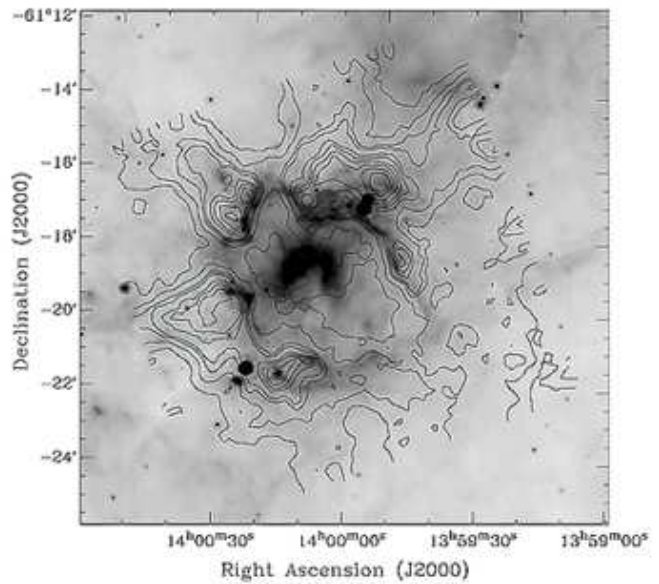}
  \caption{Isocontours of the $^{12}$CO emission integrated between $-42$
  and $-52$~km.s$^{-1}$ superimposed on the 24 $\mu$m emission of RCW~82 taken from
  the Spitzer-MIPSGAL survey.}
  \label{Fig.MIPS24_iso12CO}
\end{figure}

$\bullet$ The molecular component centred at $-48$~km~s$^{-1}$
almost completely surrounds the ionized region. It consists of a
thin layer (about 50$\arcsec$ thick (beam-corrected) or 0.8~pc at
a distance of 3.4~kpc) together with a number of large clumps.
Some of these clumps have cometary tails. H$\alpha$ emission from
RCW~82 is observed at a velocity of about $-50$~km~s$^{-1}$
(Russeil et al.~\cite{rus98}). Based on velocity and morphological
arguments we associate the molecular component centred at about
$-48$~km~s$^{-1}$ with the \HII\ region. This point is well
illustrated by Fig.~\ref{Fig.MIPS24_iso12CO} which shows the
excellent correspondence between molecular emission integrated
between $-42$ and $-52$ km~s$^{-1}$ and the border of the \HII\
region (outlined by the 24~$\mu$m emission of the PDR).

$\bullet$ The two components centred at $-39$ and
$-35$~km~s$^{-1}$ are probably not associated to RCW~82. They
mainly consist of diffuse emission that extends throughout the
field. The $-39$~km~s$^{-1}$ component shows two emission peaks in
the middle of the field, in the direction of the interior of
RCW~82 (Fig.~\ref{Fig.pannel_lignedevise_12CO}). The
$-35$~km~s$^{-1}$ component shows four emission peaks. Three of
them are observed in the direction of the interior of RCW~82. They
do not correspond to any absorption structures observed for
example in the optical; thus these condensations are probably
situated behind RCW~82 (the kinematic distances associated with
the $-35$~km~s$^{-1}$ velocity are 8.54 and 2.60~kpc, using the
galactic rotation curve of Brand \& Blitz~\cite{bra93}).

$\bullet$ The $+16$~km~s$^{-1}$ component lies very far from the
\HII\ region, at 12.6~kpc from the sun, according to the Galactic
rotation curve of Brand \& Blitz (\cite{bra93}).

To conclude, based on velocity and morphology arguments, we
consider in the following that the structure in the velocity range
$-42$ to $-52$~km~s$^{-1}$ is associated with RCW~82, but we bear
in mind that the molecular emission centred at $-55$~km~s$^{-1}$
could also be in interaction with RCW~82, even if it seems to
be morphologically unrelated.\\

\begin{figure*}
 \includegraphics[width=155mm]{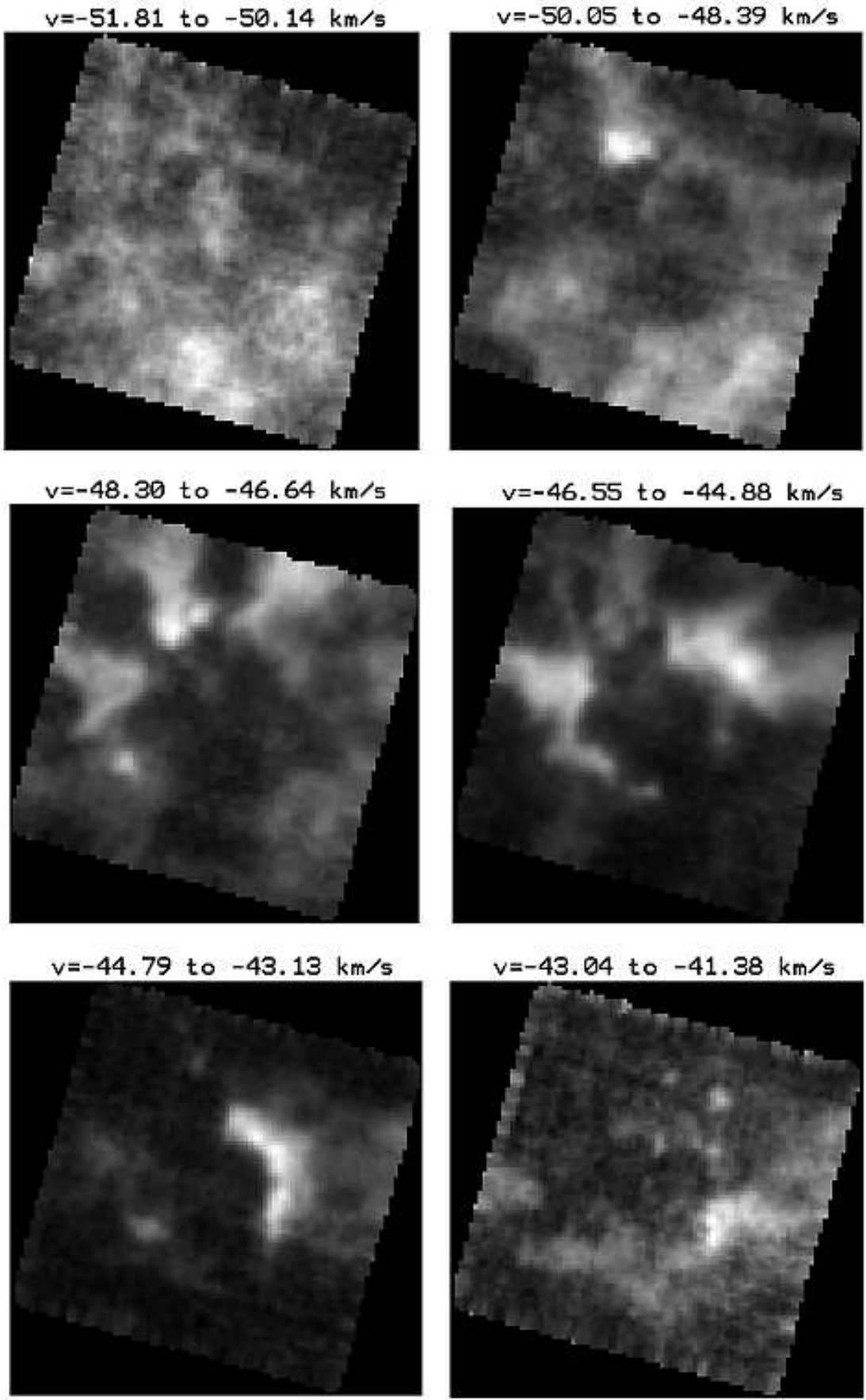}
  \caption{Integrated $^{12}$CO emission  in the direction of RCW~82.
  Velocity ranges are indicated on the top of each image.}
  \label{Pannel_12CO}
\end{figure*}
\begin{figure*}
 \includegraphics[width=155mm]{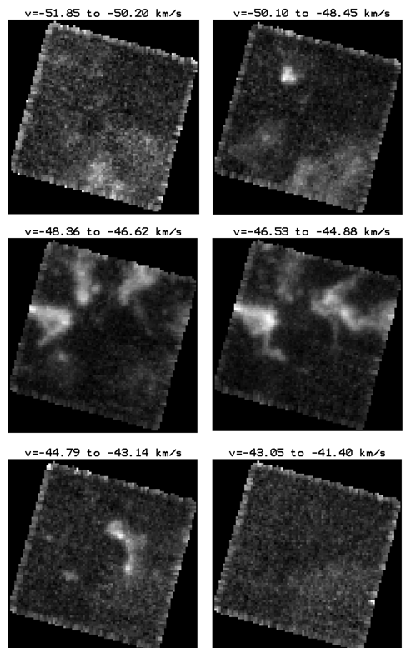}
  \caption{Integrated $^{13}$CO emission in the direction of RCW~82.}
  \label{Pannel_13CO}
\end{figure*}

Fig.~\ref{Pannel_12CO} and \ref{Pannel_13CO} show $^{12}$CO and
$^{13}$CO emission associated with RCW~82, integrated over small
velocity ranges of $\sim$1.6~km~s$^{-1}$ wide, from $-51.81$ to
$-41.38$~km~s$^{-1}$. The emission peaks around $-48$~km~s$^{-1}$.
This velocity is slightly different from the velocity of the
ionized gas: $-50$~km~s$^{-1}$ from the H$\alpha$ line (Russeil et
al.~\cite{rus98}) or $-51$~km~s$^{-1}$ from the H~109$\alpha$ line
(Caswell \& Haynes \cite{cas87}).\\

We now briefly describe the molecular material associated with
RCW~82 shown in figures \ref{Pannel_12CO} and \ref{Pannel_13CO}.
To make this description clearer, we use the numbering of the
molecular structures shown in Fig.~\ref{Fig.NH2_via_13CO}.

From $-51.85$ to $-50.14$~km~s$^{-1}$ two clumps of $^{12}$CO
emission are seen southeast and southwest of the \HII\ region;
these two clumps are also seen in $^{13}$CO. Some faint $^{12}$CO
emission is observed towards the centre of the ionized region but
is not detected in $^{13}$CO.

From $-50.05$ to $-48.40$~km~s$^{-1}$ the southern emission is
still present, and the clump $\#$1a appears north of the \HII\
region. Also, clump $\#$3 is seen on the southeast border.

From $-48.30$ to $-46.63$~km~s$^{-1}$ the emission is seen mainly
on the northeastern part of the ionized region. Clump $\#$3
observed on the east border is brighter. The same is observed for
structure $\#$1 which now shows a cometary tail and a secondary
peak on the west (structure 1b). Structures $\#$2, $\#$4 and $\#$5
appear on either side of the ionized region (north and northeast
of the \HII\ region). They show cometary tails extending
perpendicular to the ionization front. These structures are
separated by holes in the CO emission.
Fig.~\ref{fig.Halpha_NH2via13CO} shows ionized gas (H$\alpha$
emission) passing between structures $\#$1 and $\#$2. The
`leakage' of FUV radiation into these gaps might help to form the
shapes of the molecular condensations.

From $-46.55$ to $-43.13$~km~s$^{-1}$: Structure $\#$1 disappears,
but structures $\#$5 and $\#$6 become brighter. On this same side
we also see a thin layer of material adjacent to the ionization
front (structure $\#$7). This is probably the shell of collected
material compressed between the ionization front and the shock
front expanding with the \HII\ region, as predicted by the collect
and collapse theory (Elmegreen \& Lada~\cite{elm77}). The western
part of this layer is not exactly at the same velocity but begins
to appear. On the western counterpart the fragment is now larger
and denser and seems to be elongated at both its extremities. This
strange shape is compatible with the idea that some structures are
due to pre-existing clouds.

From $-44.79$ to $-43.14$~km~s$^{-1}$ almost all the clumps have
disappeared. In this velocity range we see the western part of the
collected material (structure $\#$7) which seems to be about
50$\arcsec$ in size (beam corrected) on the $^{12}$CO emission map
(Fig.~\ref{Pannel_12CO}). The $^{13}$CO emission
(Fig.~\ref{Pannel_13CO}) shows that the layer seems to be
fragmented. The collect and collapse process predicts
gravitational instabilities in the collected layer. The
observation of cores agrees with this point, showing that the
fragmentation phase has occurred for RCW~82. This point is
discussed in section~\ref{sect.Models}.

Only the brightest condensations are detected by our C$^{18}$O
observations (not shown here), mainly the clumps situated north of
RCW~82 (structures $\#$1a, $\#$2, $\#$5 and $\#$6). This confirms
that the surrounding molecular material is probably denser in the
north than in the south.

\subsubsection{H$_2$ column density}

We estimated the $N$(H$_2$) column density from the $^{12}$CO and
$^{13}$CO observations; the C$^{18}$O emission is too weak over
the field to be used, except in some specific locations. An upper
limit to the total column density of $^{13}$CO can be derived by
making the assumption that all rotational levels are thermalized
with the same excitation temperature $T_{\rm{K}}=T_{\rm{ex}}$ (LTE
assumption). The map of the excitation temperature is
obtained from the peak brightness temperature of $^{12}$CO(1-0),
assumed to be optically thick, and corrected for the cosmic
background emission, via

\begin{equation} T_{\rm{ex}}=\frac{5.532}{\ln(\frac{5.532}{T_{B}^{*}+0.840}+1)}. \end{equation}

\noindent The map we obtain shows temperatures up to 37~K.

The total column density map of $^{13}$CO is obtained assuming
that the $^{13}$CO(1-0) emission is optically thin, so that

\begin{equation} N_{\rm{Tot}}(^{13}\rm{CO})=U_{T_{\rm{ex}}}\frac{8\pi}{c^3}\frac{\nu_0^2}{g_{u}A_{ul}}e^{\frac{E_{l}}{kT_{\rm{ex}}}}\frac{k}{h}\int_{-\infty}^{+\infty}T_{\rm{mb}}\,d\rm{v} \label{eq_Ntot}
\end{equation}

\noindent where $U_{T_{\rm{ex}}}=0.7597 \times T_{\rm{ex}}$
is the partition function for $^{13}$CO (M\"uller et
al.~\cite{mul05}).

Then, assuming LTE we use the conversion factor
H$_2$/$^{13}$CO=6$\times$10$^5$ (obtained from the ``canonical
value" $^{12}$CO/H$_2$=10$^{-4}$ (van Dishoeck et
al.~\cite{vanDi92}) and the isotopic abundance
$^{12}$CO/$^{13}$CO=60), we can calculate the H$_2$ column
density:

\begin{equation} N(\rm{H_2})= 6 \times 10^{5}~N(^{13}\rm{CO}) \label{eq.NH2via13CO} \end{equation}

\begin{figure}
 \includegraphics[width=90mm ]{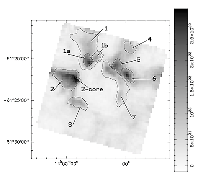}
  \caption{Map of H$_2$ column density obtained from the $^{13}$CO data,
  assuming that the $^{13}$CO emission is optically thin. The structures
  discussed in the text are identified. The solid contours are the limits chosen to
  define the structures presented in Table~\ref{tab:MassesCond}.}
  \label{Fig.NH2_via_13CO}
\end{figure}

\noindent The column density map of H$_{2}$ obtained using
$^{13}$CO data via equations \ref{eq_Ntot} and
\ref{eq.NH2via13CO} is presented in
Fig.~\ref{Fig.NH2_via_13CO}. It shows several condensations as
well as structures resulting from the accumulated material
around the ionized region during its expansion.

It is important to note that the coefficients used for the
conversion from $N$($^{12}$CO) and $N$($^{13}$CO) to $N$(H$_{2}$)
are quite uncertain, and therefore the masses of the condensations
are uncertain as well.

\subsubsection{Molecular condensations}

\begin{table*} \caption{Mass of the molecular structures observed around RCW~82.\label{tab:MassesCond}}
\begin{tabular}{c c c c c c c}
\hline \hline
Structure & $\alpha_{2000}$ & $\delta_{2000}$ & velocity range & N(H$_2$) peak & N(H$_2$) level & mass \\
 &  &  & (km~s$^{-1}$) &  (cm$^{-2}$) & (cm$^{-2}$) & (\msol) \\
 \hline
 1 & 13$^{\rm h}$ 59$^{\rm m}$ 36.86$^{\rm s}$ & -61$\degr$ 19$\arcmin$ 14.5$\arcsec$ & $-50.10$ to $-45.80$ &  & 2.5 $\times$ 10$^{21}$ & 2300 \\
 1a & 13$^{\rm h}$ 59$^{\rm m}$ 37.08$^{\rm s}$ & -61$\degr$ 20$\arcmin$ 18.97$\arcsec$ & $-50.10$ to $-45.80$ & 2.4 $\times$ 10$^{22}$ & 1.3 $\times$ 10$^{22}$ & 620 \\
 1b & 13$^{\rm h}$ 59$^{\rm m}$ 30.40$^{\rm s}$ & -61$\degr$ 19$\arcmin$ 55.00$\arcsec$ & $-48.45$ to $-45.98$ & 1.7 $\times$ 10$^{22}$ & 3.0 $\times$ 10$^{21}$ & 187 \\
 2 & 13$^{\rm h}$ 59$^{\rm m}$ 57.96$^{\rm s}$ & -61$\degr$ 22$\arcmin$ 17.3$\arcsec$ & $-48.09$ to $-44.79$ &  & 9.0 $\times$ 10$^{21}$ & 2560 \\
 2-core & 13$^{\rm h}$ 59$^{\rm m}$ 52.12$^{\rm s}$ & -61$\degr$ 22$\arcmin$ 30.84$\arcsec$ & $-48.09$ to $-44.79$ & 2.7 $\times$ 10$^{22}$ & 2.3 $\times$ 10$^{22}$ & 355 \\
 3 & 13$^{\rm h}$ 59$^{\rm m}$ 47.14$^{\rm s}$ & -61$\degr$ 25$\arcmin$ 18.90$\arcsec$ & $-46.35$ to $-44.51$ & 7.9 $\times$ 10$^{21}$ & 1.5 $\times$ 10$^{21}$ & 255 \\
 4 & 13$^{\rm h}$ 58$^{\rm m}$ 57.15$^{\rm s}$ & -61$\degr$ 18$\arcmin$ 45.22$\arcsec$ & $-48.09$ to $-46.35$ & 8.7 $\times$ 10$^{21}$ & 6.5 $\times$ 10$^{21}$ & 283 \\
 5 & 13$^{\rm h}$ 59$^{\rm m}$ 10.39$^{\rm s}$ & -61$\degr$ 20$\arcmin$ 30.95$\arcsec$ & $-46.90$ to $-45.34$ & 1.5 $\times$ 10$^{22}$ & 9.0 $\times$ 10$^{21}$ & 386 \\
 6 & 13$^{\rm h}$ 58$^{\rm m}$ 58.78$^{\rm s}$ & -61$\degr$ 21$\arcmin$ 56.65$\arcsec$ & $-46.71$ to $-44.79$ & 2.1 $\times$ 10$^{22}$ & 9.0 $\times$ 10$^{21}$ & 626 \\
 7 & 13$^{\rm h}$ 59$^{\rm m}$ 5.36$^{\rm s}$ & -61$\degr$ 22$\arcmin$ 54.91$\arcsec$ & $-46.35$ to $-43.14$ & 1.4 $\times$ 10$^{22}$ & 2.0 $\times$ 10$^{21}$ & 1100 \\
 8 & 14$^{\rm h}$ 00$^{\rm m}$ 01.75$^{\rm s}$ & -61$\degr$ 24$\arcmin$ 01.1$\arcsec$ & $-52.79$ to $-56.61$ &  & 1.5 $\times$ 10$^{22}$ & 1500 \\
 8-core & 13$^{\rm h}$ 59$^{\rm m}$ 59.61$^{\rm s}$ & -61$\degr$ 24$\arcmin$ 19.1$\arcsec$ & $-52.79$ to $-56.61$ & 3.24 $\times$ 10$^{22}$ & 2.0 $\times$ 10$^{22}$ & 780 \\
 \hline
\end{tabular}
\end{table*}

We identified several molecular structures around RCW~82 (see
Figs.~\ref{Fig.NH2_via_13CO} and \ref{Fig.struct-55}). These
structures are clumps and fragments, selected by eye in the
$^{12}$CO and $^{13}$CO data cubes. Each structure is observed
over a given velocity range. Maps of the column density were
obtained for each of these velocity ranges. The mass of the
structures is then deduced from the column density integrated over
the spatial extent of the structures. The results are given in
Table~\ref{tab:MassesCond}. The first column gives the
identification of the structure, as in Fig.~\ref{Fig.NH2_via_13CO}
and \ref{Fig.struct-55}. Columns 2 and 3 are the coordinates of
the approximate centr of the structure. Column 4 is the velocity
range over which the structure is seen, and column 5 gives the
column density at the emission peak. Column 6 is the N(H$_{2}$)
value used to define the integration boundary and derive the mass,
and column 7 gives the resulting masses. The masses given in
Table~1 have been derived from N($^{13}$CO).

We compared these masses to those obtained using the
$^{12}$CO emission alone, via eq.~17 of Rosolowsky et al.
(\cite{ros06}). This formula uses a CO-to-H$_2$ conversion factor
$X_{\rm{CO}}$=2$\times$10$^{20}$~cm$^{-2}$~(K~km~s$^{-1}$)$^{-1}$
and takes into account the presence of helium in the mass
calculation. The derived masses differ by up to 50 percent from
the values given in Table~\ref{tab:MassesCond}. This discrepancy
can be explained by the uncertainty on the conversion factors we
used, and by the fact that the densest clumps are probably
optically thick in $^{12}$CO (thus the $^{12}$CO masses are
underestimated). This argument is confirmed by the N(H$_2$) column
density map we obtain via the $^{12}$CO (not shown here), which
shows lower column densities in the direction of the densest part
of the condensations than those obtained with the $^{13}$CO.

Massive molecular structures are thus observed all around RCW~82.
They contain enough material to form massive objects, stars or
clusters. A number of these structures are adjacent to the
ionization front ($\#$ 1a, $\#$ 1b, $\#$ 2-core). It is not
possible to know if they were pre-existent clumps suddenly
submitted to the pressure of the ionized gas, or if they are
fragments resulting from the gravitational collapse of the
collected shell. Structures $\#$ 3 and $\#$ 7 are clearly parts of
this collected shell. Figs.~\ref{Pannel_12CO} and
\ref{Pannel_13CO} show that substructures are present in the
collected layer.

\subsection{A search for YSOs towards RCW~82\label{sect.study_YSOs}}

\subsubsection{The global search}

Our purpose is to study star formation in the vicinity of RCW~82
by detecting all the YSOs around RCW~82 and looking at their
position with respect to the ionized gas and to the molecular
condensations. Has star formation been triggered by the expanding
\HII\ region? For that, we will use 2MASS data, supplemented by
NTT observations in two fields, Spitzer-GLIMPSE and MIPS data. We
do not know if all the sources seen in the direction of RCW~82 lie
at the same distance as the \HII\ region and are associated with
it. However, the high density of YSOs located in the immediate
surroundings of RCW~82 indicates that the association is highly
probable.

\begin{figure}
 \includegraphics[width=90mm ]{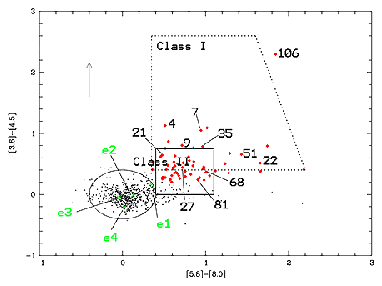}
  \caption{Colour-colour diagram for Spitzer-GLIMPSE sources in the direction of RCW~82.
  The ellipse shows the location of main-sequence and giant stars, the rectangle the
  Class~II objects, and the dotted quadrilateral the area of the Class~I objects, according to Allen et al. (\cite{all04}).
  The arrow shows the reddening  shift corresponding to an extinction of $A_{V}=40$~mag.}
  \label{cc_Spitzer}
\end{figure}

Allen et al.~(\cite{all04}) shown that YSOs have specific
mid-IR colours depending on their masses and their evolutionary
stages. As a first step we based our YSO selection on the
$[3.6]-[4.5]$ versus $[5.8]-[8.0]$ colour-colour diagram, obtained
from magnitudes given in the Spitzer-GLIMPSE catalogue. For our
study we considered an area of 12$\farcm$5 $\times$12$\farcm$5
centred on RCW~82, which corresponds to the field covered by the
Mopra observations. This colour-colour diagram is presented in
Fig.~\ref{cc_Spitzer}. We selected sources that satisfy the
criteria $[3.6]-[4.5]>0.2$ and $[5.8]-[8.0]>0.4$. With these
colour criteria we should select all the Class I objects,
intermediate Class I/II objects (i.e. sources lying in the
overlap region between Class I and Class II) and most of the
Class II objects. We omit stars with $[3.6]-[4.5]$ between 0 and
0.2  in order to avoid reddened giant stars. This selection leads
to a list of 58 YSO candidates which are presented as red symbols
in Fig.~\ref{cc_Spitzer}. Some sources discussed in the text are
identified in this figure.

In a second step, we used the MIPS-24~$\mu$m image. We measured
the magnitudes of MIPS sources using DAOPHOT (PSF photometry). As
most sources are observed in the direction of bright emission
filaments, PSF photometry gives better results than aperture
photometry. Many MIPS-24~$\mu$m sources are YSOs detected in our
previous selection. We checked with the $J-H$ vs. $H-K$ and
$[3.6]-[4.5]$ vs. $[5.8]-[8.0]$ diagrams and found that some other
24~$\mu$m sources are giant stars; they will not be considered
hereafter. Only four more MIPS-24~$\mu$m sources could be young
stars (stars $\#$98, $\#$100, $\#$101, $\#$103); they were not
selected before because at least one magnitude was missing in the
GLIMPSE catalogue.

We added a source to our list of YSOs: source $\#$106 (identified
in Fig~\ref{fig_M1I4I2}) is a very bright IR star, apparently
isolated. It is not measured in the Spitzer-GLIMPSE catalogue, and
we used DAOPHOT to estimate its magnitudes on the post BCD
Spitzer-GLIMPSE images. This star is saturated at 24~$\mu$m. The
other strong IR source seen close to it, source $\#$107, is a
giant star. To summarize, we found 63 sources which are possibly
YSOs associated with RCW~82.

For all these sources we used (when available) the 2MASS data
(Skrutskie et al.~\cite{skr06}), in order to complete the
wavelength coverage and to obtain the spectral energy
distributions (SEDs) of these objects. Table~\ref{tableEtoiles}
gives the list of the sources studied in this paper. The first
column gives the source designation; the second and third columns
give the corresponding coordinates. Columns 4 to 6 are the 2MASS
or NTT magnitudes, columns 7 to 10 are the Spitzer-GLIMPSE
magnitudes, and column 11 gives the MIPS-24~$\mu$m magnitude.
Sources $\#$51 and $\#$106 are saturated on the MIPS 24~$\mu$m
map; we have used short exposure AKARI maps at 15 and 24~$\mu$m to
measure their flux (see Zavagno et al. in preparation for a
description of these data). We obtained respectively 1.97~Jy at
15~$\mu$m and 6.716~Jy at 24~$\mu$m for source $\#$51, and 3.06~Jy
(15~$\mu$m) and 8.13~Jy (24~$\mu$m) for source $\#$106.

\begin{figure}
 \includegraphics[width=90mm ]{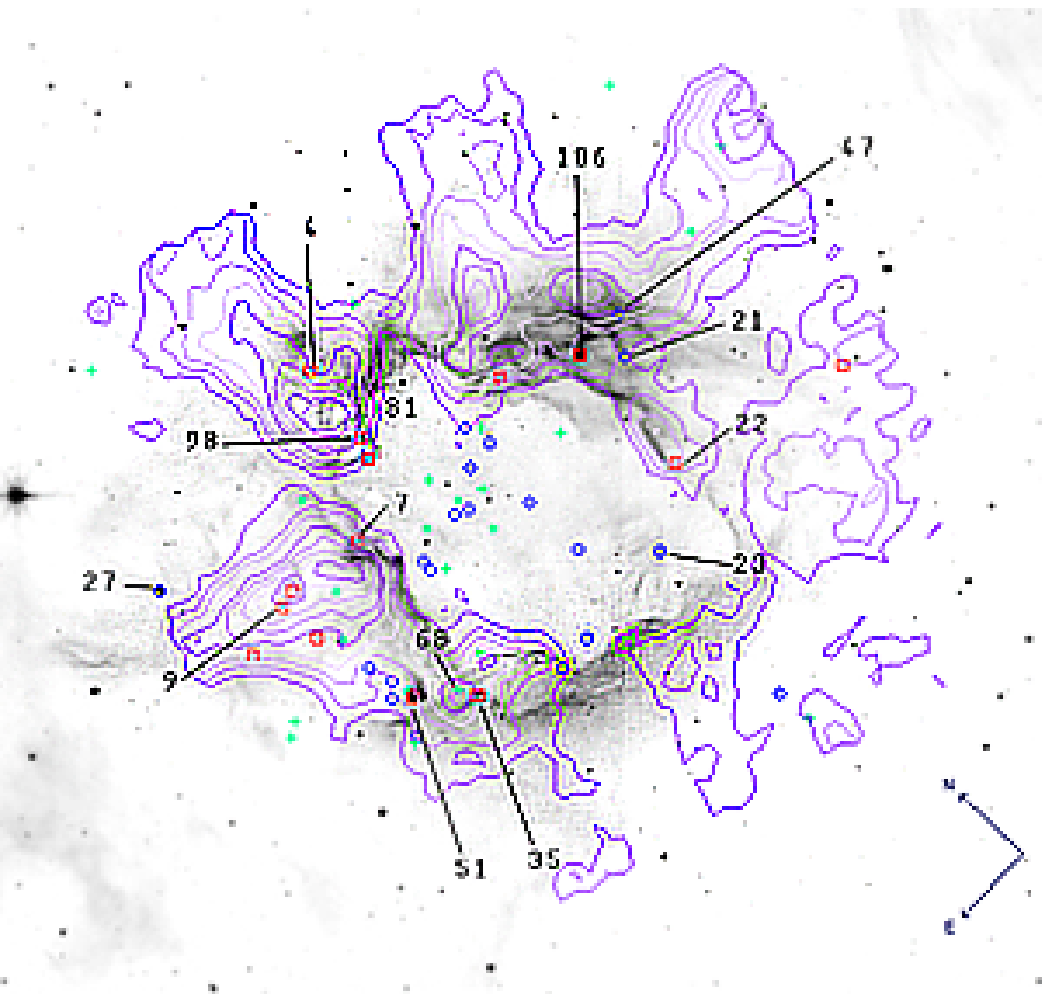}
  \caption{Superposition of the integrated $^{13}$CO emission contours on
  the 8~$\mu$m image. The selected YSOs are shown. Red squares, blue circles
  and green crosses respectively correspond to Class~I, Intermediate and Class~II sources. The
  sources discussed in the text are identified by their number according
  to Table~\ref{tableEtoiles}.}
  \label{fig_13CO_selection}
\end{figure}

Fig.~\ref{fig_13CO_selection} shows the spatial distribution of
the 63 YSOs candidates in the vicinity of RCW~82 compared to the
molecular structures seen between $-42$ and $-52$~km~s$^{-1}$.
Several sources are seen on the borders of the \HII\ region and
towards the ionized gas. The evolutionary stages for the 63 young
sources were estimated (Class I, Class II, intermediate Class
I/II) according to their position in $[3.6]-[4.5]$ versus
$[5.8]-[8.0]$ diagram, in the 2MASS $J-H$ vs. $H-K_{\rm{S}}$
diagram and to their SEDs.

Several sources are very luminous at mid-IR wavelengths, and have
rising SEDs. Some of them have emission detected in all bands from
1.24 to 24~$\mu$m. Stars $\#$51 and $\#$68 are members of the
field~1 cluster and are discussed in section~\ref{NTT_Cluster}.
Star $\#$7 is from field~2 of the SofI-NTT observations and is
also discussed in section \ref{NTT_Cluster}. Source $\#$106 will
be discussed in section \ref{Isolated_star}.

\begin{table*} \tiny{ \caption{JHK, Spitzer-GLIMPSE and MIPS magnitudes of discussed stars.\label{tableEtoiles}}
\begin{tabular}{c c c c c c c c c c c}
\hline\hline
Star & $\alpha_{2000}$ & $\delta_{2000}$ & $J$ & $H$ & $K_{\rm{S}}$ & [3.6] & [4.5] & [5.8] & [8.0] & [24]\\
 \hline
 \multicolumn{11}{c}{\bf Objects discussed around RCW~82} \\
 \hline
 4 & 13$^{\rm h}$ 59$^{\rm m}$ 34.76$^{\rm s}$ & $-$61$\degr$ 19$\arcmin$ 22.10$\arcsec$ & 17.586: & 15.055 & 12.408 & 10.127 & 8.999 & 8.771 & 8.258 & 5.04  \\
 9 & 14$^{\rm h}$ 00$^{\rm m}$ 01.65$^{\rm s}$ & $-$61$\degr$ 21$\arcmin$ 54.76$\arcsec$ & 15.892: & 15.058 & 12.693 & 9.703 & 8.897 & 8.136 & 7.419 & 4.15  \\
 21 & 13$^{\rm h}$ 59$^{\rm m}$ 01.26$^{\rm s}$ & $-$61$\degr$ 22$\arcmin$ 59.87$\arcsec$ & 16.652 & 14.821 & 13.304 & 11.178 & 10.534 & 10.020 & 9.544 & ... \\
 22 & 13$^{\rm h}$ 59$^{\rm m}$ 06.76$^{\rm s}$ & $-$61$\degr$ 24$\arcmin$ 55.77$\arcsec$ &  &  &  & 13.305 & 12.791 & 11.198 & 9.539 & ... \\
 23 & 13$^{\rm h}$ 59$^{\rm m}$ 17.56$^{\rm s}$ & $-$61$\degr$ 25$\arcmin$ 49.26$\arcsec$ & 17.067: & 15.109 & 13.574 & 12.150 & 11.611 & 10.902 & 10.043 & 7.48 \\
 27 & 14$^{\rm h}$ 00$^{\rm m}$ 12.63$^{\rm s}$ & $-$61$\degr$ 20$\arcmin$ 13.23$\arcsec$ & 17.188: & 12.422 & 9.390 & 7.040 & 6.611 & 5.932 & 5.205 & 1.85 \\
 35 & 13$^{\rm h}$ 59$^{\rm m}$ 50.89$^{\rm s}$ & $-$61$\degr$ 25$\arcmin$ 22.23$\arcsec$ & 13.723 & 12.207 & 10.780 & 8.622 & 7.837 & 7.074 & 6.110 & 2.58 \\
 47 & 13$^{\rm h}$ 58$^{\rm m}$ 57.77$^{\rm s}$ & $-$61$\degr$ 22$\arcmin$ 25.58$\arcsec$ & 16.613 & 14.451 & 12.407 & 10.201 & 9.583 & 9.143 & 8.687 & ... \\
 81 & 13$^{\rm h}$ 59$^{\rm m}$ 31.91$^{\rm s}$ & $-$61$\degr$ 20$\arcmin$ 37.43$\arcsec$ & 17.605 & 15.261 & 13.145 & 11.306 & 11.065 & 10.543 & 9.635 & ... \\
 98 & 13$^{\rm h}$ 59$^{\rm m}$ 36.76$^{\rm s}$ & $-$61$\degr$ 20$\arcmin$ 48.25$\arcsec$ & 16.267: & 15.305 & 13.566 & 10.088 & 9.238 & 8.472 & ... & 4.01 \\
 106 & 13$^{\rm h}$ 59$^{\rm m}$ 05.86$^{\rm s}$ & $-$61$\degr$ 22$\arcmin$ 26.50$\arcsec$ & ... & ... &  & 10.867 & 8.570 & 6.382 & 4.485 & ... \\
 107$^{\footnotesize b}$ & 13$^{\rm h}$ 59$^{\rm m}$ 03.57$^{\rm s}$ & $-$61$\degr$ 22$\arcmin$ 15.80$\arcsec$ & 10.870 & 8.074 & 6.320 & ... & ... & ... & ... & 1.47 \\
 \hline
 \multicolumn{11}{c}{\bf Cluster, NTT field 1} \\
 \hline
 51 & 13$^{\rm h}$ 59$^{\rm m}$ 57.61$^{\rm s}$ & $-$61$\degr$ 24$\arcmin$ 36.85$\arcsec$ & 17.811$^{\footnotesize a}$ & 13.795$^{\footnotesize a}$ & 11.127$^{\footnotesize a}$ & 8.192 & 7.538 & 6.157 & 4.724 & \\
 68 & 13$^{\rm h}$ 59$^{\rm m}$ 57.48$^{\rm s}$ & $-$61$\degr$ 24$\arcmin$ 29.38$\arcsec$ & 14.968$^{\footnotesize a}$ & 12.731$^{\footnotesize a}$ & 11.294$^{\footnotesize a}$ & 9.671 & 9.304 & 9.123 & 8.116 & \\
 53 & 13$^{\rm h}$ 59$^{\rm m}$ 59.94$^{\rm s}$ & $-$61$\degr$ 24$\arcmin$ 22.14$\arcsec$ & 19.762$^{\footnotesize a}$ & 16.419$^{\footnotesize a}$ & 14.088$^{\footnotesize a}$ & 11.921 & 11.290 & 10.692 & 10.059 & \\
 36 & 13$^{\rm h}$ 59$^{\rm m}$ 58.14$^{\rm s}$ & $-$61$\degr$ 24$\arcmin$ 10.27$\arcsec$ & 17.695$^{\footnotesize a}$ & 14.973$^{\footnotesize a}$ & 13.354$^{\footnotesize a}$ & 11.353 & 10.819 & 10.245 & 9.626 & 5.64 \\
 43 & 13$^{\rm h}$ 59$^{\rm m}$ 59.16$^{\rm s}$ & $-$61$\degr$ 23$\arcmin$ 43.467$\arcsec$ & 19.191$^{\footnotesize a}$ & 16.308$^{\footnotesize a}$ & 14.772$^{\footnotesize a}$ & 13.162 & 12.693 & 12.123 & 11.275 & \\
 102$^{\footnotesize b}$ & 13$^{\rm h}$ 59$^{\rm m}$ 48.40$^{\rm s}$ & $-$61$\degr$ 24$\arcmin$ 20.23$\arcsec$ & 15.287$^{\footnotesize a}$ & 11.323$^{\footnotesize a}$ & 9.341$^{\footnotesize a}$ & 7.791 & 7.715 & 7.249 & 7.087 & 5.80 \\
 A2 & 13$^{\rm h}$ 59$^{\rm m}$ 57.40$^{\rm s}$ & $-$61$\degr$ 24$\arcmin$ 37.2$\arcsec$ & ... & 16.962$^{\footnotesize a}$ & 14.216$^{\footnotesize a}$ & ... & ... & ... & ... & ... \\
 C2 & 13$^{\rm h}$ 59$^{\rm m}$ 57.12$^{\rm s}$ & $-$61$\degr$ 24$\arcmin$ 34.4$\arcsec$ & 19.359$^{\footnotesize a}$ & 17.381$^{\footnotesize a}$ & 15.951$^{\footnotesize a}$ & ... & ... & ... & ... & ... \\
 E2 & 13$^{\rm h}$ 59$^{\rm m}$ 56.43$^{\rm s}$ & $-$61$\degr$ 24$\arcmin$ 29.2$\arcsec$ & 17.865$^{\footnotesize a}$ & 15.721$^{\footnotesize a}$ & 14.826$^{\footnotesize a}$ & ... & ... & ... & ... & ... \\
 F2 & 13$^{\rm h}$ 59$^{\rm m}$ 56.88$^{\rm s}$ & $-$61$\degr$ 24$\arcmin$ 41.2$\arcsec$ & ... & 15.594$^{\footnotesize a}$ & 13.889$^{\footnotesize a}$ & ... & ... & ... & ... & ... \\
 L2 & 13$^{\rm h}$ 59$^{\rm m}$ 57.89$^{\rm s}$ & $-$61$\degr$ 24$\arcmin$ 25.46$\arcsec$ & 18.226$^{\footnotesize a}$ & 15.828$^{\footnotesize a}$ & 14.303$^{\footnotesize a}$ & 12.128 & 11.857 & 11.514 & ... & ... \\
 M2 & 13$^{\rm h}$ 59$^{\rm m}$ 58.13$^{\rm s}$ & $-$61$\degr$ 24$\arcmin$ 27.07$\arcsec$ & ... & 18.989$^{\footnotesize a}$ & 16.201$^{\footnotesize a}$ & 13.663 & 12.917 & ... & ... & ... \\
 P2 & 14$^{\rm h}$ 00$^{\rm m}$ 01.70$^{\rm s}$ & $-$61$\degr$ 24$\arcmin$ 51.40$\arcsec$ & 12.126 & 11.7096 & 11.504 & 11.270 & 11.162 & ... & ... & ... \\
 \hline
 \multicolumn{11}{c}{\bf NTT field 2} \\
 \hline
 A3 & 13$^{\rm h}$ 59$^{\rm m}$ 44.34$^{\rm s}$ & -61$\degr$ 22$\arcmin$ 43.77$\arcsec$ & 14.099$^{\footnotesize a}$ & 13.048$^{\footnotesize a}$ & 12.630$^{\footnotesize a}$ & 12.377 & 12.603 & ... & ... & ... \\
 B3 &13$^{\rm h}$ 59$^{\rm m}$ 44.85$^{\rm s}$ & $-$61$\degr$ 22$\arcmin$ 44.4$\arcsec$ & 12.488$^{\footnotesize a}$ & 12.147$^{\footnotesize a}$ & 11.960$^{\footnotesize a}$ & ... & ... & ... & ... & ... \\
 C3 & 13$^{\rm h}$ 59$^{\rm m}$ 44.71$^{\rm s}$ & $-$61$\degr$ 22$\arcmin$ 51.1$\arcsec$   & ... & 15.172$^{\footnotesize a}$ & 13.683$^{\footnotesize a}$ & ... & ... & ... & ... & ... \\
 D3 & 13$^{\rm h}$ 59$^{\rm m}$ 45.11$^{\rm s}$ &  $-$61$\degr$ 22$\arcmin$ 46.1$\arcsec$ & 15.578$^{\footnotesize a}$ & 14.249$^{\footnotesize a}$ & 13.459$^{\footnotesize a}$ & ... & ... & ... & ... & ... \\
 7 & 13$^{\rm h}$ 59$^{\rm m}$ 47.34$^{\rm s}$ & $-$61$\degr$ 22$\arcmin$ 01.69$\arcsec$ & 18.766$^{\footnotesize a}$ & 15.847$^{\footnotesize a}$ & 14.028$^{\footnotesize a}$ & 11.284 & 10.235 & 9.111 & 8.167 & ... \\
 J3 & 13$^{\rm h}$ 59$^{\rm m}$ 46.87$^{\rm s}$ & $-$61$\degr$ 22$\arcmin$ 09.26$\arcsec$ & 15.914$^{\footnotesize a}$ & 14.573$^{\footnotesize a}$ & 13.484$^{\footnotesize a}$ & 11.882 & 11.445 & 11.381 & ... & ... \\
 N3 & 13$^{\rm h}$ 59$^{\rm m}$ 48.89$^{\rm s}$ & $-$61$\degr$ 23$\arcmin$ 03.4$\arcsec$ & 17.606$^{\footnotesize a}$ & 16.076$^{\footnotesize a}$ & 14.861$^{\footnotesize a}$ & ... & ... & ... & ... & ... \\
 O3 & 13$^{\rm h}$ 59$^{\rm m}$ 49.38$^{\rm s}$ & $-$61$\degr$ 23$\arcmin$ 04.01$\arcsec$ & ... & ... & 15.035$^{\footnotesize a}$ & 12.276 & 11.453 & 10.599 & ... & ... \\
 \hline
 \multicolumn{11}{c}{\bf candidate exciting stars} \\
 \hline
 e1 & 13$^{\rm h}$ 59$^{\rm m}$ 29.93$^{\rm s}$ & $-$61$\degr$ 23$\arcmin$ 07.78$\arcsec$ & 10.090 & 9.549 & 9.093 & 8.439 & 8.275 & 8.000 & 7.656 & ... \\
 e2 & 13$^{\rm h}$ 59$^{\rm m}$ 29.17$^{\rm s}$ & $-$61$\degr$ 23$\arcmin$ 03.02$\arcsec$ & 9.934 & 9.587 & 9.403 & 9.211 & 9.218 & 9.225 & 9.098 & ... \\
 e3 & 13$^{\rm h}$ 59$^{\rm m}$ 28.11$^{\rm s}$ & $-$61$\degr$ 22$\arcmin$ 59.12$\arcsec$ & 9.741 & 9.409 & 9.188 & 8.972 & 9.020 & 9.114 & 9.149 & ... \\
 e4 & 13$^{\rm h}$ 59$^{\rm m}$ 27.39$^{\rm s}$ & $-$61$\degr$ 23$\arcmin$ 19.34$\arcsec$ & 12.792 & 10.438 & 9.352 & 8.623 & 8.826 & 8.505 & 8.472 & ... \\
 \hline
 \label{massemm}
\end{tabular}\\
\\
{$^{\footnotesize a}$ NTT magnitude}\\
{$^{\footnotesize b}$ giant stars}}
\end{table*}

\begin{figure}
 \includegraphics[width=90mm ]{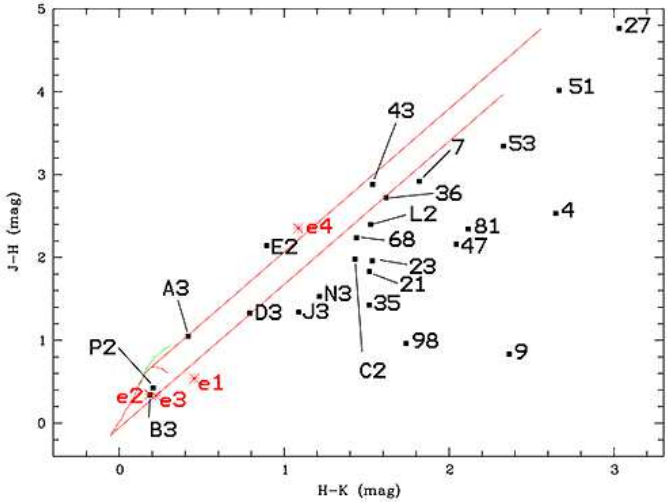}
  \caption{Colour-colour diagram with $J$, $H$ and $K_{\rm{S}}$ magnitudes (2MASS
  and NTT-SofI). The main sequence is plotted in red, and the giant branch in green.
  Reddening lines have a length corresponding to a visual extinction
  of 40~mag.}
  \label{cc_JHK}
\end{figure}

\begin{figure}
 \includegraphics[width=80mm]{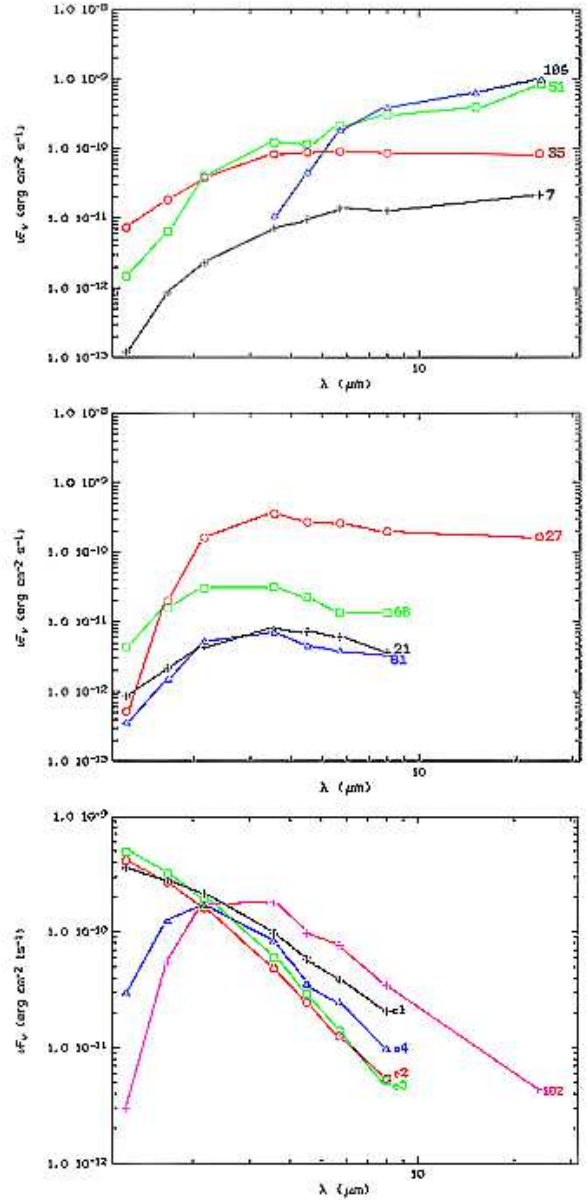}
  \caption{Spectral energy distributions of some Class~I YSOs
  (top) and Class~II and intermediate ClassI/II YSOs (middle) in the vicinity
  of RCW~82. The bottom image shows the SEDs of the four candidate exciting
  stars of the \HII\ region and that of star $\#$102, a typical giant.}
  \label{fig_SEDs_general}
\end{figure}

Several sources are located in the Class I area of the
$[3.6]-[4.5]$ versus $[5.8]-[8.0]$ diagram, tracing the emission
of the dusty envelope in which the stars are embedded. However,
Class~II objects located behind a large quantity of external dust
could be shifted into the Class~I area of the diagram. Using 2MASS
data, we constructed a $J-H$ vs. $H-K$ diagram in
Fig.~\ref{cc_JHK}. The IR-excess evident in this diagram is
generally attributed to emission from accretion disks.
Characteristic SEDs of some YSOs detected around RCW~82 are shown
in Fig.~\ref{fig_SEDs_general} (top and middle). In
Fig.~\ref{fig_SEDs_general} (top) the SEDs of sources $\#$7,
$\#$35, $\#$51 and $\#$106 are plotted. Sources $\#$4, $\#$9, and
$\#$22, identified in Fig.~\ref{fig_13CO_selection}, have similar
SEDs increasing from 1.24~$\mu$m to 24~$\mu$m. SEDs of
intermediate and Class~II YSOs, such as sources $\#$21, $\#$27
(intermediate), $\#$68 and $\#$81 (Class~II) have respectively
SEDs which become flat or slowly decrease in the mid-IR (see
Fig.~\ref{fig_SEDs_general} middle). Fig~\ref{fig_SEDs_general}
(bottom) shows, for comparison, the SEDs of the four candidate
exciting stars of the \HII\ region. Stars e2 and e3 are
main-sequence stars, and e4 is most probably a giant. On this same
graph we have plotted, for comparison with source e4, the SED of
the source $\#$102, a typical giant star.

\subsubsection{The cluster\label{NTT_Cluster}}

One of the two main star formation sites observed towards the
borders of RCW~82 is a young cluster centred at
$\alpha_{2000}$=13$^{\rm h}$ 59$^{\rm m}$ 57$\fs$4,
$\delta_{2000}$=$-$61$\degr$ 24$\arcmin$ 32$\farcs$9, on the
southeast border of RCW~82.
Fig.~\ref{NTT_Field1and2_K_I1}~(middle) presents a colour
composite image of this cluster. $K_{\rm{S} }$ (2.162~$\mu$m)
emission observed with SofI at ESO-NTT is visible in blue, and
3.6~$\mu$m Spitzer-GLIMPSE emission in red. As for the other
fields, we plotted these stars in the $J-H$ versus $H-K$
colour-colour diagram in Fig.~\ref{cc_JHK}. The sources are
identified by their number in Table~\ref{tableEtoiles}.

Several bright red sources, very luminous at near-IR wavelengths,
are seen in this NTT field. We verified, from their
magnitudes and their positions in the $J-H$ vs. $H-K$ diagram,
that they are main sequence or giant stars. The small cluster at
the centre of the field consists of several red stars. The sources
$\#$51 and $\#$68 are the most luminous in the near-IR. Their
emission dominates the emission of the cluster at longer
wavelengths. MIPS 24~$\mu$m emission is centred on these sources
which are not resolved.

Source $\#$51 presents a slight near-IR excess on the $J-H$ versus
$H-K$ diagram (Fig.~\ref{cc_JHK}) and a large extinction
($A_{V}\geq$ 40~mag). According to its position on the Spitzer
GLIMPSE $[3.6]-[4.5]$ vs. $[5.8]-[8.0]$ diagram
(Fig.~\ref{cc_Spitzer}) this source is a Class I object, whose IR
emission is dominated by an accretion envelope. The SED seen in
Fig.~\ref{fig_SEDs_general} (top) clearly increases from
1.24~$\mu$m to 8~$\mu$m. The cluster is not resolved at longer
wavelengths, and the flux density at 15~$\mu$m and 24~$\mu$m
(AKARI data) is probably dominated by the emission of $\#$51,
which is brighter and has an increasing SED up to 8~$\mu$m.

Star $\#$68 has a small near-infrared excess (Fig.~\ref{cc_JHK}).
Its SED (Fig.~\ref{fig_SEDs_general}, middle) and its position in
the Spitzer-GLIMPSE colour-colour diagram (Fig.~\ref{cc_Spitzer})
are compatible with the characteristics of a Class~II YSO, and
thus this object may have a near-IR emission dominated by an
accretion disk.

Around both these stars, SofI-NTT high resolution observations
show a population of faint red sources in the cluster. Most of
these sources have no Spitzer-GLIMPSE magnitudes because they are
not resolved, but a diffuse emission is associated with the
cluster at these wavelengths. Sources C2 and L2 appear in the
colour-colour diagram (Fig.~\ref{cc_JHK}) with a small near-IR
excess, and could be YSOs. In contrast, the colours of E2
suggest it is a giant star.

Source P2 is located southeast of the cluster and is clearly
associated with a bow-shock (see
Fig.~\ref{NTT_Field1and2_K_I1}~(middle)). According to its near-IR
magnitudes this object is a main sequence star of type $\sim$B3V
if located at the distance of RCW~82, and associated with 3.5~mag
of visual extinction. Moreover only a spectrum could confirm
its characteristics. This bow shock may originate from a wind
emitted by source P2 and interacting with material flowing away
from the cluster (more or less in the direction of the cluster).
Another possibility is that star P2 emits a wind and is moving in
the ambient medium, creating the bow shock (Povich et
al.~\cite{pov08}). Complementary data are needed to distinguish
between the two possibilities. However, the first one is favored
by the good alignment between the vertex of the bow shock and the
direction of the cluster.

Northeast of the young cluster, three other bright red sources
are seen (sources $\#$36, $\#$53 and the ``Giant"). Sources $\#$36
and $\#$53 are classified as intermediate Class I/II sources from
the colour-colour GLIMPSE diagram. This is in agreement with their
SEDs and with their positions in the $JHK$ colour-colour diagram.
The third source has the near-IR colours of a giant star.

\subsubsection{Field 2\label{NTT_Field2}}

\begin{figure}
 \includegraphics[width=75mm ]{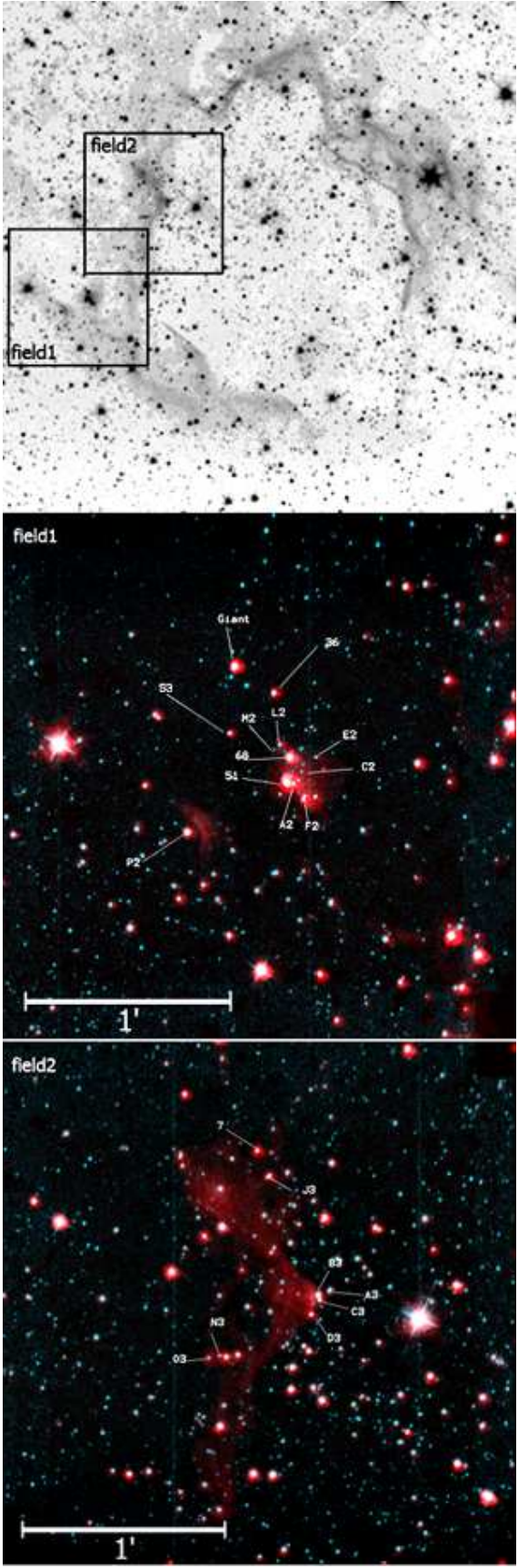}
  \caption{Top: RCW~82 at 3.6~$\mu$m, logarithmic scale. Middle and Bottom: Colour composite image of field1 and field2. NTT $K_{\rm{S}}$ emission is in blue, and
  Spitzer-GLIMPSE emission at 3.6~$\mu$m in red. The coordinates of the image
  centres are $\alpha_{2000}$=13$^{\rm h}$59$^{\rm m}$ 57$\fs$4, $\delta_{2000}$=$-$61$\degr$ 24$\arcmin$
  32$\farcs$9 (field1) and $\alpha_{2000}$=13$^{\rm h}$ 59$^{\rm m}$ 45$\fs$6, $\delta_{2000}$=$-$61$\degr$ 22$\arcmin$
  54$\farcs$5. The field sizes are 2$\farcm$6$\times$2$\farcm$6. North is to up and east is
  left.}
  \label{NTT_Field1and2_K_I1}
\end{figure}

Field 2, presented in Fig.~\ref{NTT_Field1and2_K_I1}~(bottom), is
centred on a bright border at $\alpha_{2000}$=13$^{\rm h}$
59$^{\rm m}$ 45$\fs$6, $\delta_{2000}$=$-$61$\degr$ 22$\arcmin$
54$\farcs$5. This field contains a bright filament at 8.0~$\mu$m
-- a part of the hot PDR bordering RCW~82, with a structure
pointing towards the exciting stars (Fig.~\ref{Ex_Stars}). Several
stellar objects are seen in the direction of this structure, more
or less embedded in the diffuse filamentary emission.

Only the most interesting objects are identified on
Fig.~\ref{NTT_Field1and2_K_I1}~(bottom). Most sources of the field
are main sequence or giant stars. Source C3 seems to correspond to
the tip of the structure seen at 8~$\mu$m and at 24~$\mu$m. This
source has been measured in $H$ and $K_{\rm{S}}$. It is seen in
$J$,  but not measured because it lies too close to star B3, which
is more luminous and probably a main sequence star. We used a
source of similar brightness, more isolated and measured in $J$
(17.67~mag) to estimate the position of the source C3 on the $J-H$
vs. $H-K_{\rm{S}}$ diagram; this indicates a young star. Moreover,
C3 continues to be visible at longer wavelengths, up to 24~$\mu$m,
even if no Spitzer-GLIMPSE magnitudes are available.

Other candidate YSOs are sources $\#$7, J3 and N3. Source O3 has
missing measurements in the near- and mid-IR which prevents us
from determining its nature. In addition we confirmed that the two
sources to the right of sources O3 and N3 have main sequence
near-IR colours.


\section{Discussion}

\subsection{The molecular material}

Fig.\ref{Fig.NH2_via_13CO} shows that the H$_2$ column densities
reach values up to about 3$\times$10$^{22}$~cm$^{-2}$. The four
densest condensations are seen north of RCW~82. The layer of
neutral material collected during the expansion of the \HII\
region is clearly seen in this figure.

\begin{figure}
 \includegraphics[width=90mm]{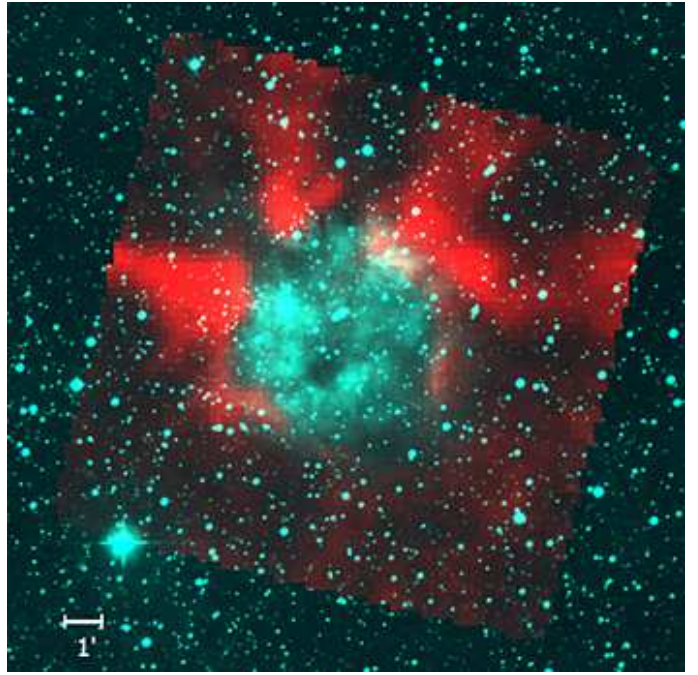}
  \caption{Colour composite image of RCW~82. H$\alpha$ emission of the
  ionized gas is seen in turquoise, and the molecular material associated
  with the H$_2$ column density map (see Fig.~\ref{Fig.NH2_via_13CO}) in red.}
  \label{fig.Halpha_NH2via13CO}
\end{figure}

Fig.~\ref{fig.Halpha_NH2via13CO} shows a composite colour image of
RCW~82 with the H$\alpha$ emission in turquoise and the integrated
$^{13}$CO emission between $-42$ and $-52$~km~s$^{-1}$ in red. It
shows that the ionized gas seems to escape from the centre of
RCW~82, flowing between condensations $\#$1 and $\#$2, and between
$\#$1 and $\#$5. We also see H$\alpha$ emission between the
collected layer of accumulated material (structure $\#$7) and the
dense condensations $\#$5 and $\#$6. This point is important
because it demonstrates that the most massive fragments seen at
the border of RCW~82 are adjacent to the \HII\ region and thus
submitted to the pressure of the ionized gas.

One of the characteristics of the molecular emission around RCW~82
is the shape of the condensations. They present a bright core
adjacent to the ionized region and extended diffuse tails almost
perpendicular to the ionization front. These structures are seen
in the $^{12}$CO and $^{13}$CO maps. We can only speculate about
the origin of these structures: i) the exciting cluster of RCW~82
formed at the meeting point of several filaments, and what we see
is the remnant of these pre-existing filaments; ii) the \HII\
region formed and evolved in a medium with pre-existing dense
condensations. The expanding \HII\ region interacted with these
condensations, collecting more material in their direction, hence
the dense cores adjacent to the ionized gas. Then the ionized gas
escaped from the central \HII\ region in directions of lower
density zones, giving a radial shape to the molecular material
hidden behind the heads of the condensations.

Also RCW~82 is seen on the border of a large bubble, to which it
is possibly linked. The molecular component observed around
$-55$~km~s$^{-1}$ has a velocity compatible with the velocity of
the H$\alpha$ emission of RCW~82 and show filaments following the
borders of the large shell. The large shell has possibly triggered
the formation of RCW~82 or is in interaction with the neutral
material surrounding RCW~82.

\subsection{Star formation around RCW~82\label{Isolated_star}}

\subsubsection{The distribution of YSOs}

\begin{figure}
 \includegraphics[width=90mm ]{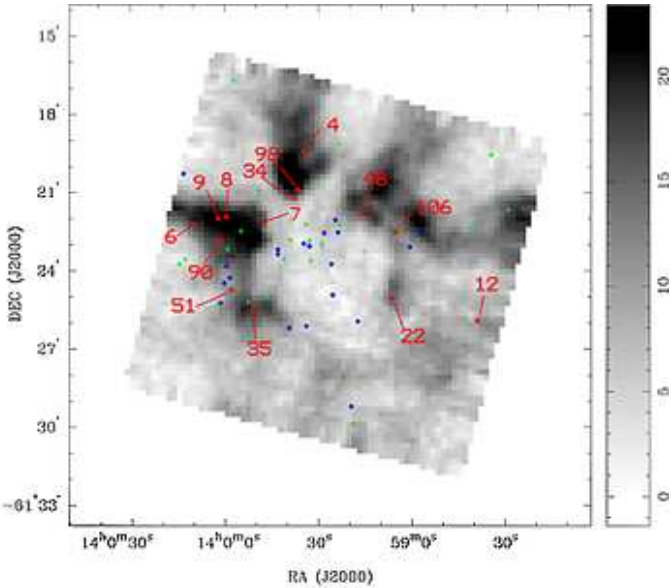}
  \caption{Map of the $^{13}$CO emission integrated between $-42$ and $-52$~km~s$^{-1}$.
  Red squares, blue circles and green crosses represent the positions of Class I,
  intermediate, and Class II YSOs, respectively.}
  \label{Fig.13CO_YSO_superposes}
\end{figure}

Fig.~\ref{Fig.13CO_YSO_superposes} shows the distribution of the
YSOs detected in the direction of RCW~82, superimposed on the
$^{13}$CO emission integrated between $-42$ and $-52$~km~s$^{-1}$.
This image shows a strong correlation between Class I sources and
the molecular emission surrounding the \HII\ region. Many of these
sources are not observed in the direction of molecular emission
peaks, but are located on the borders of the condensations.
However we have to keep in mind that some sources are possibly
linked with molecular material observed in other velocity ranges.

YSOs $\#$34 and $\#$98 are situated on the border of the molecular
structure $\#$1a, linked to RCW~82. Only molecular emission
associated with the \HII\ region (see
Fig.~\ref{Fig.pannel_lignedevise_12CO}) is observed in the
direction of these two sources, so there is a strong probability
that they are both linked to the \HII\ region.

Three other sources are very probably linked to the collected
layer of material surrounding RCW~82. These sources are $\#$48,
$\#$106 and $\#$22. They are perfectly aligned along the western
part of the collected layer (fragment $\#$7), and no molecular
emission at other velocities is seen in their direction.

Another interesting source is the Class I $\#$35. Molecular
emissions at velocities $-55$, $-35$ and $+16$~km~s$^{-1}$ are
detected in the direction of this young source, but the strongest
emission corresponds to the structure $\#$3 which is associated
with RCW~82. Thus this Class~I YSO could have been triggered by
the \HII\ region.

\subsubsection{The most massive YSOs}

We continue this discussion by focusing on the two main sites of
star formation seen on the borders of the \HII\ region.

\paragraph{Source $\#$106, an isolated YSO:}

The association of this source with RCW~82 is highly probable. It
is seen in the direction of the collected layer of materiel in
$^{12}$CO and $^{13}$CO (structure $\#$7 on
Fig.~\ref{Fig.NH2_via_13CO}), on the western side of RCW~82. No
other molecular emission is associated with source $\#$106 on the
line of sight. Thus it is highly probable that this YSO is linked
to RCW~82 and that its formation was triggered by the expansion of
the \HII\ region.

Our study in the mid-IR shows that source $\#$106 could be a Class
I candidate (Fig.~\ref{cc_Spitzer}). Furthermore this object seems
to be massive according to its SED (Fig.~\ref{fig_SEDs_general}),
and appears to be isolated. This point is confirmed by some
GEMINI 10~$\mu$m and 18.3~$\mu$m observations on which source
$\#$106 appears as a single stellar object (Zavagno et al. in
preparation).

Source $\#$106, corresponding to the MSX point source
G310.9438+00.4411, was observed by Urquhart et al.~(\cite{urq07})
at 3~cm and 6~cm. Their non-detection (at a 4$\sigma$ level of
1.1~mJy at 6~cm) allows us to derive an upper limit for the number
of ionizing photons emitted by this source, using the relation (1)
of Simpson \& Rubin (\cite{sim90}). Assuming that star $\#$106 is
probably not hot enough to ionize helium and adopting
$T_{\rm{e}}$=10000~K we obtain
$\log(N_{\rm{Lyc}}$)$\sim$45~s$^{-1}$. Thus, according to Smith et
al. (\cite{smi02}), YSO $\#$106 cannot contain a central source
hotter than a B1.5 star.

We used the online fitting tool of Robitaille et al.
(\cite{rob07}, http://caravan.astro.wisc.edu/protostars/) to model
the SED of this source. We used data from Spitzer-GLIMPSE
3.6~$\mu$m to MIPS 70~$\mu$m (this last measurement is highly
uncertain due to the underlying emission of the PDR). Most of the
models show a massive and dominant disk: considering the 50 first
best models the disk's mass is high and can reach 0.26~$\msol$;
also in most cases the envelope accretion rate is zero or very
small. Another important conclusion is that the central object is
massive, between 7 and 18~$\msol$. The inclination angle and the
extinctions (both interstellar and circumstellar) are poorly
constrained.

Two remarks can be made: i) models with important and dominant
disks are compatible with the presence of the strong silicate
absorption feature seen on the GEMINI spectra of the source
(Zavagno et al. in preparation) This absorption feature indicates
that source $\#$106 has a disk which is seen nearly edge-on; ii)
masses of more than 10~$\msol$ for the central source are
incompatible with the non-detection of radio emission at 3 and
6~cm by Urquhart et al.~(\cite{urq07}).

Thus source $\#$106 is probably a Class~II source containing a
massive central object. Whitney et al. (\cite{whi04}) shown a
trend for higher temperature sources to be redder, especially
[3.6]-[4.5], than the colours usually attributed to Class~II
sources (the box in Fig.~\ref{cc_Spitzer}); a Class~II YSO with a
hot central source can be mistaken for a Class~I source. Thus the
mass of the central source of $\#$106 and the hypothetical
presence of an accreting envelope remain to be ascertained.

\paragraph{The cluster:}

The second main site of star formation in the direction of RCW~82
is the young cluster at the eastern border of the \HII\ region.
The brightest star at mid-IR wavelengths, source $\#$51, is
centred on the MSX point source G311.0341+00.3791. This was also
observed by Urquhart et al.~(\cite{urq07}) at 3~cm and 6~cm,
but no radio continuum emission was detected. As for star $\#$106,
this means that no star massive enough to ionize hydrogen is
present in this cluster.

\begin{figure}
 \includegraphics[width=90mm]{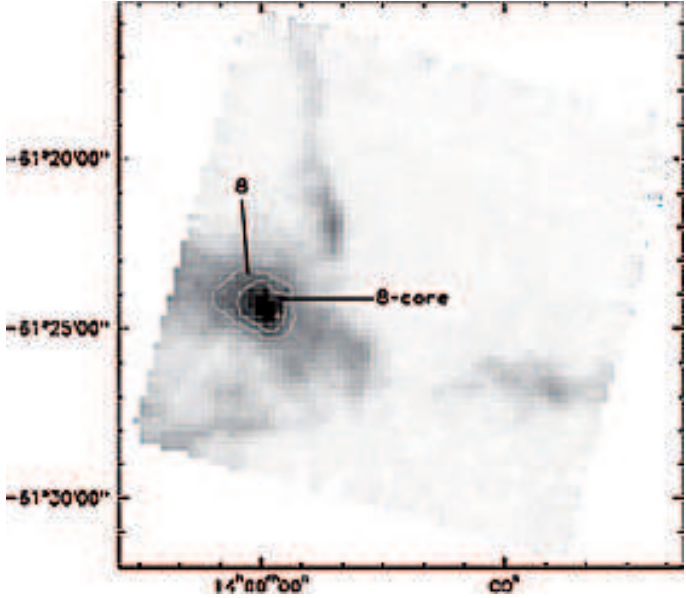}
  \caption{N(H$_2$) map of the molecular emission integrated between $-56.6$ and
  $-52.8$~km~s$^{-1}$. The contours correspond to the levels of
  1.5~$\times$~10$^{22}$~cm$^{-2}$ and 2~$\times$~10$^{22}$~cm$^{-2}$ used to compute the mass
  of the condensation (Tab.~\ref{tab:MassesCond}). The position of the Class~I $\#$51 is shown by a white cross.}
  \label{Fig.struct-55}
\end{figure}

Fig.~\ref{Fig.13CO_YSO_superposes} shows that the cluster lies in
a region of low molecular emission, slightly at the back of the
collected layer, thus rather far away from the ionization front.
Fig.~\ref{Fig.struct-55} is the calculated $N(\rm{H_2}$) map based
on the $^{13}$CO and $^{12}$CO emission integrated between $-56.6$
and $-52.8$~km~s$^{-1}$. The position of source $\#$51 is shown,
in the direction of a massive condensation. Thus the young cluster
may be associated with this molecular structure. In conclusion,
the association of the cluster with RCW~82 is uncertain, depending
on the association of the molecular emission seen around
$-55$~km~s$^{-1}$ with RCW~82.

\subsection{Comparisons with models \label{sect.Models}}

In the following we estimate the age of RCW~82, to see if the
collect \& collapse process is possibly at work around this
region. We will assume that RCW~82 formed and evolved in a
homogeneous medium of density $n_0$.

If we use an ionizing photon flux of 9.0~$\times$~10$^{48}$~s$^{-1}$
(or 4.6~$\times$~10$^{48}$~s$^{-1}$ for
the exciting star (Sect.~2.2), a present radius of 3~pc for the
\HII\ region, and assume $n_0$=10$^3$~atoms~cm$^{-3}$ (this value
will be justified a posteriori), we estimate a present electron
density for the \HII\ region n$_e\sim$120~cm$^{-3}$
(90~cm$^{-3}$), a mass M(\HII)=310~$\msol$
(230~$\msol$), an expansion velocity $\sim$3.7~km~s$^{-1}$
(3.2~km~s$^{-1}$), and an age~$\sim$0.4~Myr
(0.5~Myr).

The sphere of 3~pc radius initially contained 2600$\msol$ of
material. This material is now to be found in the ionized region
and in the shell of material accumulated around the ionized gas;
thus a mass $\sim$ 2260$\msol$ (2370$\msol$) for this
shell, in rather good agreement with the total mass
($\sim$2520$\msol$) of the structures adjacent to the IF ($\#$3,
$\#$2-core, $\#$1a, $\#$1b and $\#$7; these structures are
considered as parts of the collected layer); hence the
justification of the adopted value for n$_0$.

We can now, according to Whitworth et al.~(\cite{whi94}, section
5), estimate when the fragmentation of this collected layer should
occur. Assuming $a_{\rm{S}}$=0.2~km~s$^{-1}$ for the turbulent
velocity in the collected layer, we find that the fragmentation of
the \HII\ region occurs 1.6~Myr after its formation, when its
radius is 5.74~pc. The fragmentation will be later, and for larger
radii, for higher values of $a_{\rm{S}}$. We conclude that RCW~82
is not old enough for star formation to take place in the
collected layer via the collect \& collapse process.

The formation of the YSOs present on the border of RCW~82 most
probably results from some other process of star formation like
small-scale Jeans gravitational instabilities in the collected
layer, or interactions of the IF with pre-existing condensations
(Lefloch \& Lazareff~\cite{lef94}).


\section{Conclusions\label{conclu}}

We have presented a multi-wavelength study of the \HII\ region
RCW~82. Molecular observations in $^{12}$CO and $^{13}$CO obtained
with Mopra and the near-to-mid IR study of YSOs in the direction
of this star forming region allow us to make the following
conclusions:

$\bullet$ RCW~82 is seen on the border of a large shell observed
at 8~$\mu$m, which probably surrounds an \HII\ region.
Unfortunately the spatial coverage of our molecular observations
is not sufficient to prove the link between the two regions;

$\bullet$ We have identified four candidate exciting stars of
RCW~82. Among these sources, stars $\#$e2 and $\#$e3 are probably
O-type stars responsible for the ionization of RCW~82;

$\bullet$ Molecular material is associated with RCW~82. We observe
a molecular layer around the \HII\ region, corresponding to
interstellar material accumulated around the ionized gas during
the expansion of RCW~82. We also observe elongated structures,
oriented almost perpendicular to the ionization front. We suggest
that these structures correspond to pre-existing condensations,
possibly shaped by UV radiation. Some condensations have
masses greater than 2000~$\msol$;

$\bullet$ Many YSOs are observed at the periphery of RCW~82, among
which is a non-negligible number of potential Class~I sources.
Their distribution is not uniform and most of them are observed on
the borders of RCW~82, close to the IF. It indicates that
triggered star formation is at work around this \HII\ region;

$\bullet$ The presence of a collected layer of material around
RCW~82 shows that the collect part of the collect and collapse
process has occurred. However, our age estimate of RCW~82 shows
that the \HII\ region is probably not evolved enough for the
collected shell to be in the fragmentation phase. This means that
the star formation activity observed around RCW~82 is due to other
processes;

$\bullet$ Among the YSOs we have identified a candidate isolated
massive young object, source $\#$106; its evolutionary stage
remains to be more precisely defined. The formation of this object
may have been triggered by RCW~82, as it is observed in the
direction of the shell of material collected around the \HII\
region. However this point remains to be confirmed as the
estimated age of the \HII\ region does not allow enough time for
the collapse process to occur. On the other hand, the young
cluster, observed around source $\#$51, could be associated with a
massive molecular component observed around $-55$~km~s$^{-1}$, for
which the association with RCW~82 is uncertain.


\begin{acknowledgements}

This research has made use of the Simbad astronomical database
operated at the CDS, Strasbourg, France, and of the interactive
sky atlas Aladin (Bonnarel et al.~\cite{bon00}). We thank James
Urquhart and Martin Cohen for useful discussions about the nature
of the radio sources associated with RCW~82, and the referee, E.
Churchwell, for useful comments that helped to improve the clarity
of the paper. This publication used data products from the Two
Micron All Sky Survey, the NASA/IPAC Infrared Science Archive,
which is operated by the Jet Propulsion Laboratory, California
Institute of Technology, under contract with the National
Aeronautics and Space Administration. We also used the SuperCOSMOS
H$\alpha$ survey. This work is based in part on GLIMPSE and
MIPSGAL data obtained with the Spitzer Space Telescope, which is
operated by the Jet Propulsion Laboratory, California Institute of
Technology, under NASA contract 1407.

\end{acknowledgements}

\end{document}